\documentclass[%
 reprint, superscriptaddress,
 amsmath,amssymb, aps, prb,floatfix,
]{revtex4-1}
\usepackage{booktabs}
\usepackage{color}
\usepackage{graphicx}% Include figure files
\usepackage{dcolumn}% Align table columns on decimal point
\usepackage{bm}% bold math
\usepackage{dcolumn}
\usepackage{gensymb}
\usepackage{siunitx}
\usepackage{etoolbox}
\usepackage{booktabs}
\usepackage{multirow}
\usepackage[table,xcdraw]{xcolor}
\usepackage{tabularx}
\usepackage{makecell}
\usepackage{lipsum}
\usepackage{amsmath}
\usepackage{footnote}
\usepackage{scalerel,stackengine}
\stackMath
\newcommand\reallywidehat[1]{%
\savestack{\tmpbox}{\stretchto{%
  \scaleto{%
    \scalerel*[\widthof{\ensuremath{#1}}]{\kern-.6pt\bigwedge\kern-.6pt}%
    {\rule[-\textheight/2]{1ex}{\textheight}}%WIDTH-LIMITED BIG WEDGE
  }{\textheight}% 
}{0.5ex}}%
\stackon[1pt]{#1}{\tmpbox}%
}
\makesavenoteenv{table}
\newcommand{\RNum}[1]{\uppercase\expandafter{\romannumeral #1\relax}}

\begin{document}

\title{Alchemical and structural distribution based representation for improved QML} %\title{Fast machine learning models of electronic and energetic properties consistently reach approximation errors better than DFT accuracy}

\author{Felix A. Faber}
\affiliation{Institute of Physical Chemistry and National Center for Computational
Design and Discovery of Novel Materials, Department of Chemistry, University of Basel, Switzerland.}
\author{Anders S. Christensen}
\affiliation{Institute of Physical Chemistry and National Center for Computational
Design and Discovery of Novel Materials, Department of Chemistry, University of Basel, Switzerland.}
\author{Bing Huang}
\affiliation{Institute of Physical Chemistry and National Center for Computational
Design and Discovery of Novel Materials, Department of Chemistry, University of Basel, Switzerland.}
\author{O. Anatole von Lilienfeld}
\email{anatole.vonlilienfeld@unibas.ch}
\affiliation{Institute of Physical Chemistry and National Center for Computational
Design and Discovery of Novel Materials, Department of Chemistry, University of Basel, Switzerland.}

\date{\today}% It is always \today, today,

\begin{abstract}

We introduce a representation of any atom in any chemical environment for the generation of efficient 
quantum machine learning (QML) models of common electronic ground-state properties.
The representation is based on scaled distribution functions explicitly accounting for elemental and structural degrees of freedom. 
Resulting QML models afford very favorable learning curves for properties of out-of-sample systems including 
organic molecules, non-covalently bonded protein side-chains, (H$_2$O)$_{40}$-clusters, as well as diverse crystals.
The elemental components help to lower the learning curves, and, through interpolation across the periodic table,
even enable ``alchemical extrapolation'' to covalent bonding between elements not part of training, as evinced for 
single, double, and triple bonds among main-group elements. 
%Another key difference between our newly developed representation and most existing representations is that it can simultaneously measure conformational  {\it and} compositional distances between compounds very efficiently, using Gaussian smearing over elements. This makes it well suited for compositionally diverse sets.
\end{abstract}

\maketitle

\section{Introduction}

Ground-state properties of chemical compounds can generally be estimated with acceptable accuracy using methods such as \textit{ab initio} quantum chemistry or density functional theory (DFT)~\cite{JensenCompChem}. 
However, these can be computationally expensive and therefore have a limited applicability, especially 
for larger systems.
Alternatively, inductive quantum machine learning (QML) models
can infer properties directly, or even predict the electron density which in turn can be used to calculate properties~\cite{brockherde2017bypassing}, 
by training on a large data sets of reference property/compound pairs. 
ML models can have an exceptional trade off between predictive accuracy and computational cost.
For example, in 2017 we showed that QML models can estimate hybrid DFT atomization energies, as well as
several other properties, of medium sized organic molecules with prediction errors lower than chemical accuracy ($\sim$0.04 eV)---multiple orders of magnitude faster than hybrid DFT~\cite{faber2017fast}. 

% Kernel ridge regression (KRR)~\cite{muller2001introduction, scholkopf2002learning, Vovk2013, kernel_ridge_regression2}, for example,
% estimates a property $p$ of a compound $\mathbf{C}$ using a weighted sum of kernels as basis functions $p^{est}(\mathbf{C}) = \sum_i^N  \alpha_i K(\mathbf{C},\mathbf{C}^{\mathrm{train}}_i)$.  
% $K(\mathbf{C},\mathbf{C}^{\mathrm{train}}_i)$ is the kernel and the sum runs over all $N$ number of compounds $\mathbf{C}^{\mathrm{train}}_i$ in the training set. $\alpha_i$ are the weights, 
% which are obtained through linear regression. 
The system variables defining the ground-state properties
of a given compound are its external potential, a simple function of interatomic distances and nuclear charges. 
However, when using this information directly to measure similarity results in QML models with rather disappointing predictive power. 
This can be mitigated by transformation of system variables into ``representations''. 
Such transformations can either be designed by human intuition, 
or be included in the learning problem, e.g.~when using neural networks (NN)
which include representation learning in the supervised learning task. 
Letting a NN find the representation has proven to yield models with low out-of-sample prediction
errors~\cite{DeepTensorNN_2017, NeuralMessagePassing,duvenaud2015convolutional}.
This approach, however, has the drawback that representation and model are intermingled within the NN,
making it less amenable to human understanding, interpretation, adaptation, and further improvement. 
Furthermore, such machine designed representations do not necessarily lead to better QML performance
than human design based representations (vide infra).

% {\color{red}
% INCLUDE DISCUSSION ON
% -what defines system/what defines property  within QML philosophy
% -briefly refer to TensorNN and message passing NN which (successfully) include the representation learning task in the supervised learning task: Drawback: Obfuscating the model, less amenable to human understanding and interpretation ... also, not necessarily better performing than a human based design of the representation (vide infra).
% }

% Many representations are based on features which are known to correlate with the target property,
% or contain information about the compound, such as~\cite{Meredig2014Screening} and ~\cite{Ward2017Voronoi}.
% The ML model then finds out which features are relevant redundant,
% and tries to map the descriptor to the target property.
% These models can in many cases preform very well on the studied class of compounds. 
% However, they are, by construction, system and property dependent
% and requires {\it a-priori} knowledge of the system. 
% An alternative way of representing compounds, is instead to more
% explicitly encode 3D structure and chemical elements information
% about the system, using for example interatomic distances and angles. 
There are many ways of manually encoding the 3D structure and chemical composition of a compound
into a suitable representation. For example, we can represent a compound 
as a list of interatomic potentials~\cite{ML0,bob,BAML}. 
Another approach consists of creating a fingerprint of the compound, 
transforming internal coordinates into a fixed set of numbers.
For example, this can be done by projecting the coordinates on to a 
set of basis functions~\cite{Behler2007Symmetry},
or by creating a ``fingerprint'' from the topology of the structure~\cite{rogers2010extended}.
Distributions of internal coordinates represent another systematic approach, 
shown to yield well performing QML models applicable throughout chemical space~\cite{anatole_qc2013,Muller_Gross_crystal}.
Additional use of bags containing angular and dihedral distributions has led to further
improvements in resulting QML models~\cite{BAML,amons,faber2017fast,huo2017unified}.
%More specifically, these representations contain a 1D distribution 
%of the interatomic distances for all element-pairs,
%and a 1D distribution of the internal angels
%for all element-triplets. 
%An important shortcoming of these representations is that they typically
%decouple angular from distance information, i.e the representation does not (directly)
%specify which distances and angles belong to the same atomic centers.
Bagging based on atom types, however, severely hinders resulting QML models from transferring what has been learned from one atom type to another---a desirable feature for chemically diverse systems.
% which will deteriorate its
% performance for chemically diverse systems.

\begin{figure}
 \includegraphics[width=0.85\linewidth]{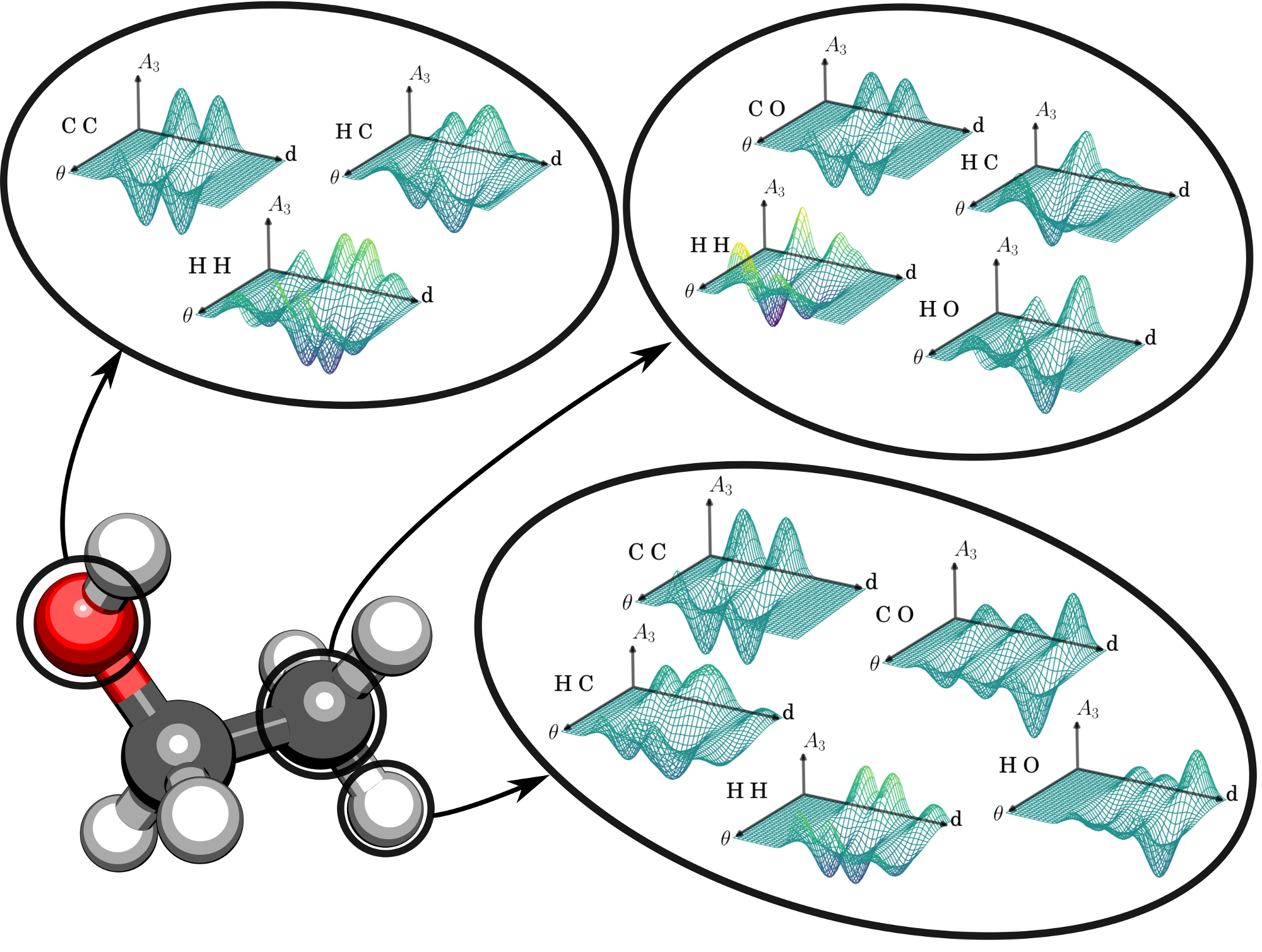}
 \caption{ \label{fig:ethanol}
  The three-body term ($A_3(\cdot)$) as a function of radial ($d$)
  and angular ($\theta$) degrees of freedom in the atomic environments
  of O, C and H (circled) in ethanol. 
  For simplicity, we show the three-body term without elemental smearing
  %($\sigma_P$, $\sigma_G \rightarrow 0$)
  where it reduces to a number
  of two-dimensional distributions for each element triplet.}
\end{figure}

In this work we introduce a new atomic environment representation,
with two key differences to previous distribution-based work. 
(i) 
The representation is not binned by atomic types.
Instead, compositional information is encoded directly into the distributions.
This allows measuring not only structural differences, but also ``alchemical''
differences between elements in the atomic environments. 
The idea of computational alchemy, amounting to continuous interpolation of Hamiltonians of two different systems, 
is well established in quantum chemistry and statistical mechanics and can be exploited for
virtual exploration campaigns in chemical space with increased efficiency~\cite{TowardsCompoundDesign2014}. 
Recently, it has been shown that alchemical estimates of covalent bond potentials 
can even surpass generalized gradient approximated DFT accuracy~\cite{Samuel2016}.
The foundation of a continuous chemical space has been reviewed previously~\cite{anatole_qc2013}.
Alchemical distance measures in the context of QML were already exploited previously 
when using the Coulomb matrix~\cite{ML0},
Fourier series distribution based representations~\cite{OAvL_FRD},
the Faber, Lindmaa, Lilienfeld, Armiento (FLLA) crystal representation~\cite{faber2016},
and within smooth overlap of atomic potentials (SOAP) representations~\cite{SOAP_apl}.
For this work we have identified a new functional form with improved performance due to alchemical
contributions to the distance measure. 
(ii) We use a set of multidimensional distributions of interatomic many-body expansions scaled by simple powerlaws,
rather than several 1D bins of internal coordinates.
The distributions are built recursively, so that an $m$-body distribution contains the same 
information as the $(m-1)$-body distribution plus additional $m$-body information. 
This particular combination combines similarity to the potential energy target function
and compliance with many known (translational, rotational, permutational) invariances.

\section{Theory}
In this section, we first motivate the ideas which have led to this study. 
Thereafter, we discuss the functional form and the variational degrees of freedom which we have introduced,
as well as the resulting compound distances.
Then, an analysis of the functional form is performed using the molecule water as an example.
Finally, numerical results for parameters optimization runs are discussed. 

\subsection{Kernel ridge regression}
In order to profit from robustness, ease of error convergence, computational efficiency, and simplicity, 
we base our studies preferably on kernel ridge regression (KRR) models~\cite{muller2001introduction, scholkopf2002learning, Vovk2013, kernel_ridge_regression2}.
However, we consider this rather a  question of taste, and believe that other regressors, such as neural networks,
will produce similar results if properly converged.

KRR estimates property $p$ of query compound $\mathbf{C}$ 
as a weighted sum of kernel basis functions placed on each of $N$ training compounds $\{{\bf C}_k\}$, 
\begin{eqnarray}
p^{est}(\mathbf{C}) & =& \sum_{k=1}^N  \alpha_k K(\mathbf{C},\mathbf{C}_k),\label{eq:KRR}\\
{\bm \alpha} & = & ({\bf K} + \lambda {\bf I})^{-1} {\bf p}^{\rm train} \label{eq:train}
\end{eqnarray}
where the solution for the weights $\{\alpha_k\}$ are obtained through linear regression with regularizer $\lambda$ 
(typically negligibly small because of absence of noise in quantum training data). 

Note that throughout this work we rely on atomistic (scalable) Gaussian kernels, 
$K({\bf C}, {\bf C}') = \sum_{I \in {\bf C}} \sum_{J \in {\bf C}'}  k(\Delta(\mathcal{A}_M(I),\mathcal{A}_M(J)))$, 
as already used in~\cite{mathias2015kernel,barker2016localized,Bartok2015GAP,amons}. 
As such, KRR renders the selection of a functional form which represents an atom in its chemical environment mandatory. 
Obviously, this choice is fundamentally related to our understanding of chemistry, and is known to dramatically
affect the performance of resulting QML models, see e.g.~\cite{BAML,faber2017fast}.
It is for this reason that we draw our inspiration from the fundamental laws of quantum mechanics 
which specify the definition of system (Hamiltonian) and property (Observable), 
and which spell out the numerical recipe which links the two~\cite{JensenCompChem}.

The genesis of this study is due to the fact that the total potential energy, the expectation value
of a compound's electronic Hamiltonian, constitutes the central figure of merit for convergence towards the
wavefunction by virtue of the variational principle. 
When considering Eq.~(\ref{eq:train}) it should be obvious that kernel (and thereby representation) are independent
of the specific property, units and property dependence are introduced through the regression weights only.
This has also already been demonstrated numerically for multiple properties using the same kernel~\cite{sk}.
As such, the role of the kernel is reminiscent of the wavefunction which can be used to predict arbitrarily 
many observables by evaluating the expectation values of the corresponding operators, always using the same wavefunction: 
Once the kernel is inverted, arbitrarily many sets of regression coefficients can easily be generated
provided that their corresponding property reference values have been provided. 
The potential electronic energy being the central property in quantum mechanics which defines the wavefunction 
it is therefore plausible to assume that a representation, optimized for energy predictions only, is
fundamentally more advantageous than representations obtained by minimizing prediction errors of alternative observables.
Consequently, the focus on this study has been to identify a representation which is inspired by 
the energy changes occurring due to changes in chemical composition and covalent and non-covalent bonding.
The accuracy quantum mechanics when predicting other properties (observables) as expectation values of operators depends crucially
on the quality of the wavefunction. Here, we follow a similar argument: 
The better the representation the better the energy prediction, 
implying that energy prediction errors can be minimized in the functional space of the representation, 
enabling systematic convergence towards an ideal representation.

\subsection{Representation}
We use a set of interatomic $M$-body expansions $\mathcal{A}_M(I) = \{A_{1}(I),A_{2}(I),A_{3}(I), \dots , A_{M}(I)\}$ 
 which contain up to $M$-body interactions to represent the structural and
 chemical environment of an atom $I$ in compound $\mathbf{C}$.
$A_m(I)$ is a weighted sum that runs over all $m$-body interactions. 
Each element in the sums consists of Gaussian basis functions, 
placed on structural and elemental degrees of freedom,
and multiplied by a scaling function $\xi_m$. 
Structural values encode geometrical information about the system,
such as interatomic distances or angles. 
As elemental parameters we use the period $P$ and group $G$ from the periodic table.
The scaling functions $\xi_m$ are used to weigh the importance of each Gaussian, 
based on internal system coordinates. 
We now consider only the first three distributions in $\mathcal{A}_M(I)$ for an atom $I$. 
We have also derived, implemented and tested the 4-body $A_4(I)$ distributions.
We refer to the supporting information (SI) for the derivation.
The predictive accuracy improvements of resulting QML models, however, were found to be negligible in comparison to the $3$-body expansion.
As such, we 
Also, the computational cost for generating large kernel matrices increases substantially 
when going from third to fourth order terms.

The first-order expansion $A_1(I)$ accounts for chemical composition (stoichiometry)
and is modeled by a Gaussian function placed on period $P_I$ and group $G_I$ of element $I$:
\begin{eqnarray}
A_1(I) &= & \mathcal{N}(\mathbf{x}^{(1)}_I) \; = \; e^{-\frac{(P_I-\chi_1)^2}{2\sigma_P^2}-\frac{(G_I-\chi_2)^2}{2\sigma_G^2}} 
\end{eqnarray}
where $\mathbf{x}^{(1)}_I=\{P_I,\sigma_P;G_I,\sigma_G \}$, with respective widths $\sigma_P$ and $\sigma_G$.
$\sigma_P$ and $\sigma_G$ can be seen as elemental smearing parameters,
which control the near-sightedness of elements in the periodic table.
$\chi_1$ and $\chi_2$ represent dummy variables for period and group, to be integrated out when 
evaluating the Euclidean distance (see Eq.~(\ref{eq:L2_distance})).
For $A_1(I)$, the scaling function is set to unity, since stoichiometry is geometry independent. 
We are not aware of other representations in the literature which employ 
similar distribution functions in the periodic table.

$A_2(I)$ is a product of $A_1(I)$ and a sum that runs over all neighboring atoms $i$:
$A_2(I)=\mathcal{N}(\mathbf{x}^{(1)}_I)\sum_{i \neq I} \mathcal{N}(\mathbf{x}^{(2)}_{iI}) \xi_2(d_{i I})$,
$\mathbf{x}^{(2)}_{iI}=\{d_{i I},\sigma_d;P_i,\sigma_P;G_i,\sigma_G\}$, where
 $d_{iI}$ and $\sigma_d$ correspond to the interatomic distance at which
 a Gaussian is placed, and its width, respectively. 
$\xi_2$ corresponds to the 2-body, interatomic distance dependent, scaling function which
takes the form of the power laws discussed below.
Note that letting $\sigma_P$ and $\sigma_G$ approach zero
is equivalent to using a radial distribution function (RDF) for each element pair.
This attribute of the representation holds for any of $A_m(I)$.
I.e., $\sigma_P,\sigma_G \rightarrow 0$ is equivalent to creating a separate distribution for each chemical element $m-$tuple in $A_m(I)$.
While atom pair-wise distribution functions are rampant as representation choice, especially for fitting potential energy surfaces
of systems with fixed chemical composition, to the best of our knowledge, combining them with scaling functions is novel.

$A_3(I)$ is the logical extension from $A_2(I)$,
it has a different scaling function with an additional summation,
running over all neighboring atoms $j$:
$A_3(I)=\mathcal{N}(\mathbf{x}^{(1)})\sum_{i \neq I} \mathcal{N}(\mathbf{x}^{(2)}_{iI})\sum_{j \neq i,I} \mathcal{N}(\mathbf{x}^{(3)}_{ijI}) \xi_3(d_{i I},d_{j I},\theta_{i j}^I)$,
$\mathbf{x}^{(3)}_{ijI}=\{\theta_{i j}^I,\sigma_{\theta};P_j,\sigma_P;G_j,\sigma_G\}$.
$P_j$ and $G_j$, similarly to $P_i$ and $G_i$,
corresponds to the period and group of atom $j$.
Again, $\xi_3(d_{i I},d_{j I},\theta_{i j}^I)$ is the (three-body) scaling function, 
and $\theta_{i j}^I$ the principal angle
between the two distance vectors $\vec{r}_{Ii}$ and $\vec{r}_{Ij}$
which span from $I$ to $i$ and $I$ to $j$, respectively.
$\sigma_\theta$ is the width of the Gaussian placed at $\theta_{ij}^I$.
Letting $\sigma_d$ go to infinity in $A_3$ is equivalent to using 
a type of angular distribution function (ADF), which in one form or another has already
been used
in several representations~\cite{amons,faber2017fast,huo2017unified}.
$A_3$ can therefore be seen as a generalized ADF containing more structural information.
Fig.~\ref{fig:ethanol} illustrates how $A_{3}(I)$ looks for a hydrogen, carbon, and the oxygen atom in ethanol.
Three-body distributions are less frequent as representation choice, 
and again, to the best of our knowledge, combining them with scaling functions is novel.

The scaling functions $\xi$ we have chosen for this work correspond to simple power laws. They have been modified from the leading order two- and three-body dispersion laws
by London, $1/r^6$, and Axilrod-Teller-Muto~\cite{atm,atm2}, $1/r^9$. Such dispersion expressions were previously already used by some of us~\cite{amons}.
Our scaling functions, however, use different exponents for the radial decay, 
and set the $C_6$ and $C_9$ coefficients to unity, as early tests indicated better performance for this choice.
For periodic systems, however, a very large cutoff radius
would be needed in order to converge the distances 
between two atomic environments, when using the optimized exponents.
We have therefore augmented the scaling functions by a previously used soft cutoff function~\cite{christensen2015improving}, 
which goes to zero at 9 \AA.

\subsection{Distances and scalar products}
In order to train and evaluate the  KRR model in Eq.~(\ref{eq:KRR}), proper distance measures must be specified.
We have found good performance when using as a distance between two atomic environments 
$\mathcal{A}_M(I)$ and $\mathcal{A}_M(J)$ a weighted sum of the distances between each $m$-body expansion:
$\Delta(\mathcal{A}_M(I),\mathcal{A}_M(J))^2 \equiv  \sum^{M}_{m = 0}\beta_m \Delta(A_{m}(I),A_{m}(J))^2$. 
Here, $\beta_m$ is another hyperparameter, which weighs the importance of each expansion order. 

The distances between each distribution term are evaluated as Euclidean ($L_2$) norms, as shown in Eq.\ref{eq:L2_distance}.
$\varsigma_m$ are normalization factors, which  ensures that all individual basis functions integrates to $1$ in the $L_2$-norm. All integrals can be solved analytically since they consist of a sum of Gaussian products.  The explicit form of the integrals can be found in SI.

\begin{widetext}

\begin{equation}
\begin{aligned}
\label{eq:L2_distance}
\Delta(A_m(I),A_m(J))^2 =  &  \dfrac{1}{\varsigma_m^2} \int_{\mathbb{R}^{3m - 1}}d\chi_1 \dotsi d\chi_{3m - 1} (A_m(I) - A_m(J))^2 
\end{aligned}
\end{equation}

\begin{equation}
\begin{aligned}
  \dfrac{1}{\varsigma_1^2}\int_{\mathbb{R}^{2}} d\chi_1 d\chi_2 A_{1}(I) A_{1}(J)  =& \dfrac{1}{2} \exp(-\dfrac{(P_I - P_J)^2}{4\sigma_P^2}-\dfrac{(G_I - G_J)^2}{4\sigma_G^2}) \\
 \dfrac{1}{\varsigma_2^2}\int_{\mathbb{R}^{5}} d\chi_1 \dotsi d\chi_5 A_{2}(I) A_{2}(J)  =& \dfrac{1}{2\,\sqrt[]{2}} \exp(-\dfrac{(P_I - P_J)^2}{4\sigma_P^2}-\dfrac{(P_I - P_J)^2}{4\sigma_G^2}) \\
 \sum^{n_I}_{i \neq I} \xi_2(d_{i I}) \sum^{n_J}_{j \neq J} & \exp(-\frac{(d_{j J} - d_{i I})^2}{4 \sigma_d^2}-\dfrac{(P_i - P_j)^2}{4\sigma_P^2}-\dfrac{(G_i - G_j)^2}{4\sigma_G^2})\xi_2(d_{j J})  \\
 \dfrac{1}{\varsigma_3^2}\int_{\mathbb{R}^{8}}d\chi_1 \dotsi d\chi_8 A_{3}(I) A_{3}(J)  =& \dfrac{1}{16} \exp(-\dfrac{(P_I - P_J)^2}{4\sigma_P^2}-\dfrac{(G_I - G_J)^2}{4\sigma_G^2}) \nonumber\\
 \sum^{n_I}_{i \neq I} \sum^{n_J}_{j \neq J} & \exp(-\dfrac{(d_{j J} - d_{i I})^2}{4 \sigma_d^2}-\dfrac{(P_i - P_j)^2}{4\sigma_P^2}-\dfrac{(G_i - G_j)^2}{4\sigma_G^2}) \\
\sum^{n_I}_{k \neq i,I} \xi_2(d_{i I},d_{k I},\theta_{i k}^I) \sum^{n_J}_{l \neq j,J} & \exp(-\dfrac{(\theta_{i k}^I - \theta_{j l}^J)^2}{4 \sigma_{\theta}^2}-\dfrac{(P_k - P_l)^2}{4\sigma_P^2}-\dfrac{(G_k - G_l)^2}{4\sigma_G^2})\xi_3(d_{j J},d_{l J},\theta_{j k}^J)
\end{aligned}
\end{equation}

\end{widetext}

Note that third and fourth order terms become prohibitively expensive to calculate directly.
However, this can to a large extent be circumvented by slightly modifying the distributions, and solving the angular integrals in Fourier space.
Further details about the corresponding equations and derivations can also be found in the SI.

\subsection{Comparison to other distribution based representations}

 \begin{figure}
  \includegraphics[width=0.95\linewidth]{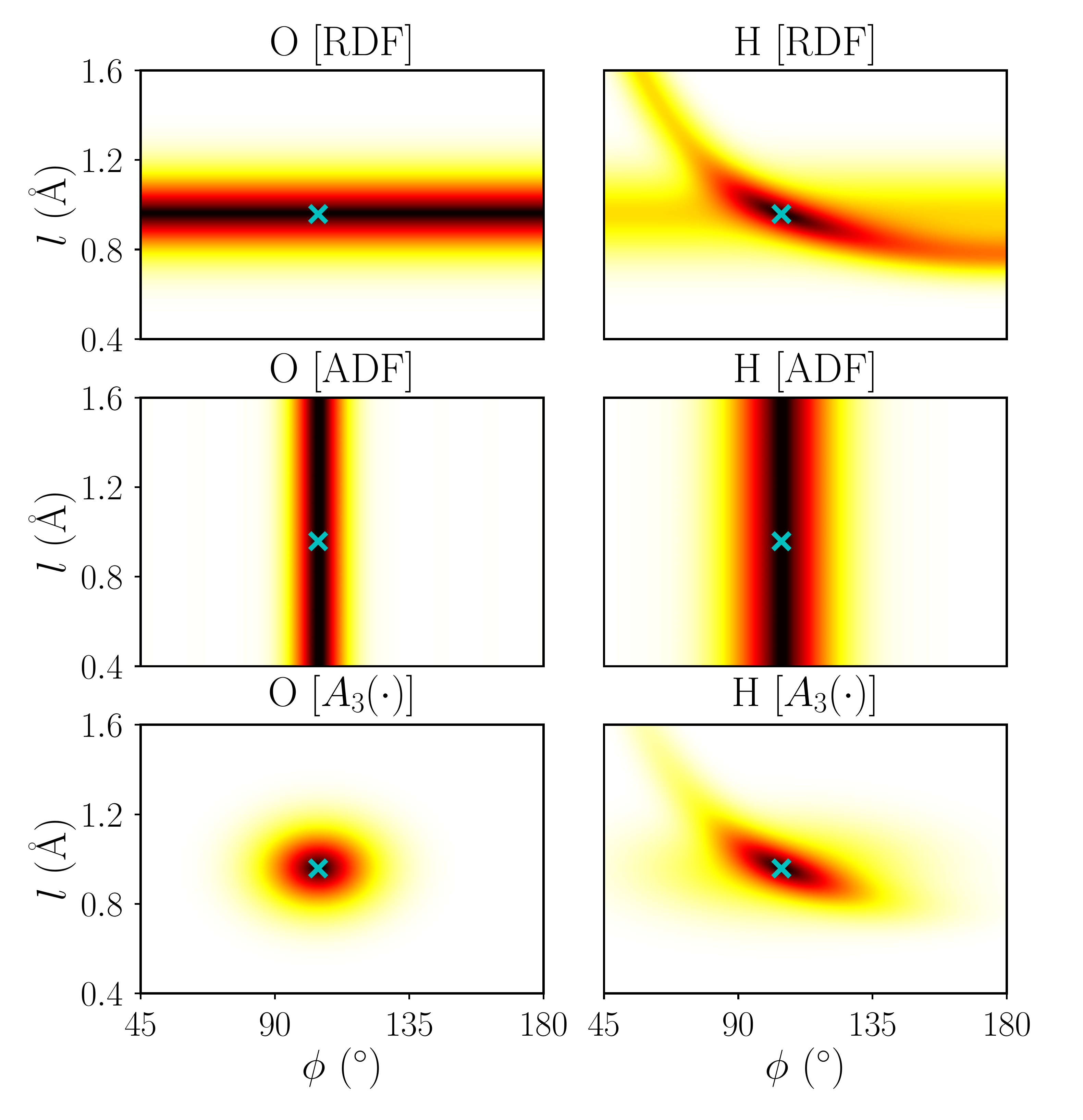}
  \caption{ \label{fig:waterScan}
     Heat maps of normalized $L_2$ distances for three representations (RDF, ADF, and our new representation). 
     The color code from black to white indicates a distance range from 0 to 1, respectively.
     The distances are measured between oxygen (LEFT) and hydrogen atoms (RIGHT) in two different water molecules.
     One water molecule is being distorted by uniform stretching of both OH bonds ($d_{\rm OH1} = d_{\rm OH2} = l$) and bending ($\phi$).
     The other water molecule is kept fixed at its equilibrium geometry (cross). 
     }
 \end{figure}

\begin{figure}
 \includegraphics[width=0.95\linewidth]{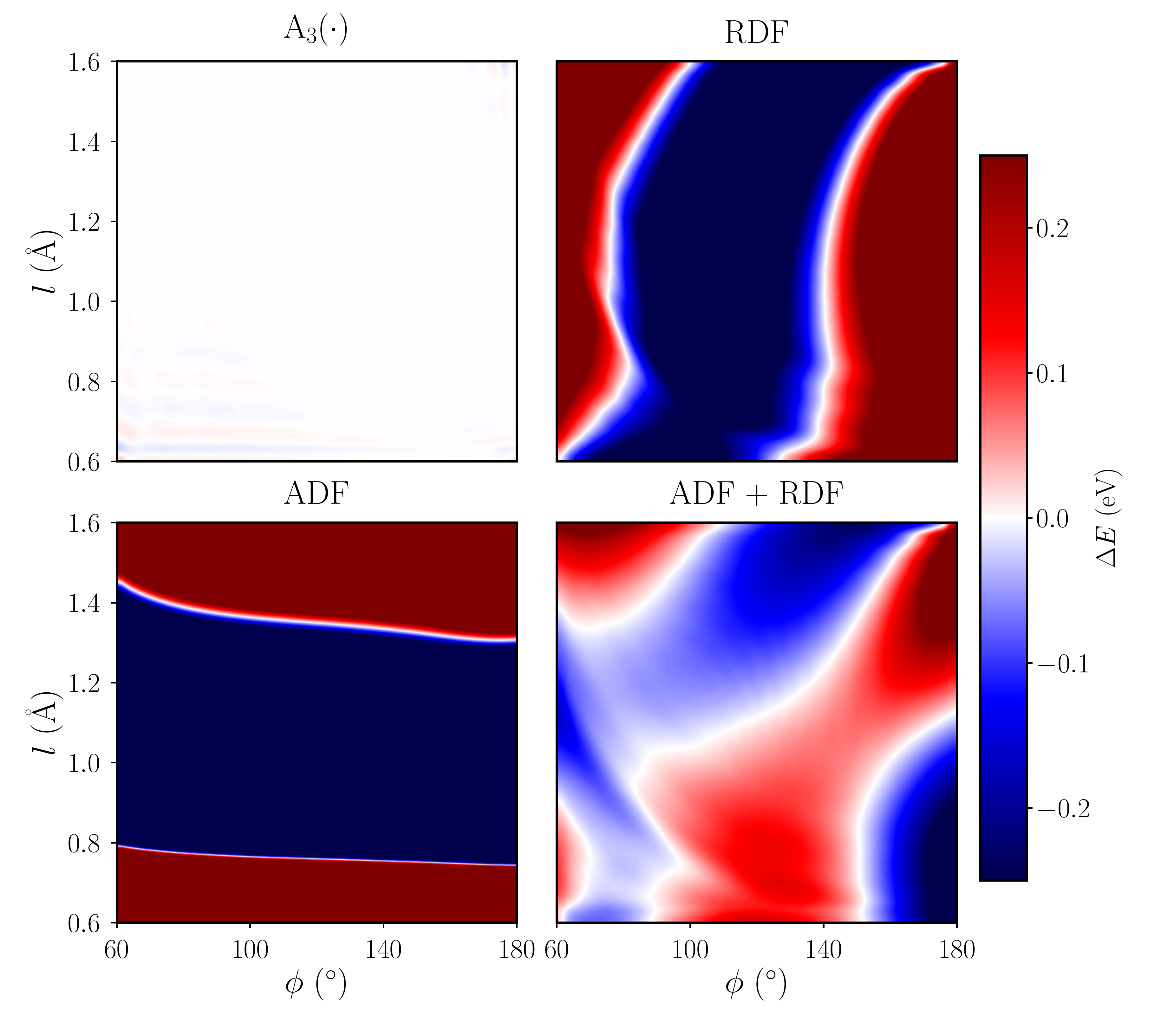}
 \caption{ \label{fig:waterError}
     Heat maps of the signed error of atomization energies in water molecule
     for the same coordinate system as in Fig.~\ref{fig:waterScan}.
     The errors correspond to linear kernels in KRR fitted to DFT calculated energies (PBE/def2svp) energies.
     Four representations have been used:
     TOP LEFT: our new $A_3(\cdot)$ (top left). 
     TOP RIGHT: radial distribution function for each element pair (RDF).
     BOTTOM LEFT: angular distribution function for each element triplet (ADF).
     BOTTOM RIGHT: RDF + ADF.
     The training data consists of a equidistant 
     grid of 50-by-50 points along $l$ and $\phi$
     within the range of the figures.
    }
\end{figure}

Probably the largest difference in how $\mathcal{A}$ represents nuclear configurations, when compared 
to many of the previously published distribution based representations, lies in the 3-body term
(since $A_2(\cdot)$ is a radial distribution functions if $\sigma_P$ and $\sigma_G$ go to zero).
In this subsection, we highlight the differences between $A_3(\cdot)$, 
or conventional ADF or RDF for representing the structure of the water molecule.

As ADF, we use $A_3(\cdot)$ with the limit $\sigma_d\rightarrow\infty$, and we model RDF  by $A_2(\cdot)$.
Furthermore, no scaling function ($\xi_2 = \xi_3 = 1$) is used and we let $\sigma_P$ and $\sigma_G$ go to zero,
since we only examine how representations distinguish structural differences among different geometries of the water molecule.
This results in $A_3(\cdot)$ and ADF being $\sum_{i \neq I} \mathcal{N}(d_{i I},\sigma_d)\sum_{j \neq i,I} \mathcal{N}(\theta_{i j}^I,\sigma_{\theta})$ and 
$\sum_{i \neq I}\sum_{j \neq i,I} \mathcal{N}(\theta_{i j}^I,\sigma_{\theta})$ for each element triplet,
and RDF being $\sum_{i \neq I} \mathcal{N}(d_{i I},\sigma_d)$ for each element pair.

Fig.~\ref{fig:waterScan} shows how the distance measure changes as one distorts the geometry
away from its equilibrium structure. Both, RDF as well as ADF result for oxygen as well as for H
in large configurational domains with substantially zero distance to the minimum, implying a severe lack of uniqueness.
$A_3$, by contrast produces a qualitatively meaningful picture with a single well 
defined well around the minimum.

We have also studied the performance for modeling the energy of the water molecule.
In Fig.~\ref{fig:waterError}, the training error for
atomization energies is shown for a linear kernel KRR model 
with $A_3(\cdot)$, ADF, RDF, or RDF + ADF as representations. 
The linear kernel is used as a difficult test in how far representations can 
model a nonlinear property, such as the energy, in terms of linear basis functions.
The errors are significantly lower when using $A_3$ instead
of the other representations, including RDF + ADF.
Generally, potential energy surfaces of a three-atom system
cannot be decomposed into many-body terms each as a function of only one internal
coordinate (internuclear distance $\mathbf{d}$ or angle $\mathbf{\theta}$).
That is, $E(\mathbf{d},\mathbf{\theta}) \neq E(\mathbf{d}) + E(\mathbf{\theta})$.
Using a ADF, RDF or a linear combination of the two however would result precisely in such a model,
as well as most force-fields.
This also explains the relatively large errors for these representations, 
as well as unreliable performance of pair-wise potentials when it comes to distorted molecules.
$A_3$ on the other hand does not decouple distances and angles,
and can, by construction, model any three-body potential.

These observations give insight as to why our new representation
performs better than the other distribution based representations:
Using ADF's and RDF's as representations might be able to capture slices of the
 many-body picture,
the fact that there is a linear mapping between $A_n(\cdot)$ and a
$n$-body potential energy surface, however, appears to make it easier to 
improve the performance also for non-linear kernels.

\subsection{Optimization}

\subsubsection{Hyper-parameters}
 The use of our representation in combination with KRR yields multiple hyperparameters. 
 While one could, in principle, attempt to optimize all of them, using  
 several data sets, and efficient optimizers, such as gradient, Monte Carlo, genetic or simplex methods,
 we have found that the problem is sensitive only to a small subset of parameters.
% This, however, proved to be challenging due to the relatively large number of hyperparameters,
% as well as the need for rather large training sets sizes in order to converge them.
 As such, the exact choice of many hyperparameters is not critical for the out-of-sample errors, 
 and resulting models perform typically well as long as values are used which have similar order of magnitude.
 Unless otherwise specified, the following hyperparameter values have been used:
$\sigma_P = \sigma_G = 1.6$,
$\sigma_{d} = 0.2$, $\sigma_{\theta} = \pi$,
%$\sigma_{\omega} = \pi$,
$\beta_1 = 1$,
$\beta_2 = \sqrt[]{8}$,
$\beta_3 = 1.6$.
For the water cluster and the SSI data set there is no to little variation in chemical composition, and no alchemical smearing has been used.

\subsubsection{Scaling powerlaw parameters}
\begin{figure}[h!]
 \includegraphics[width=0.45\linewidth]{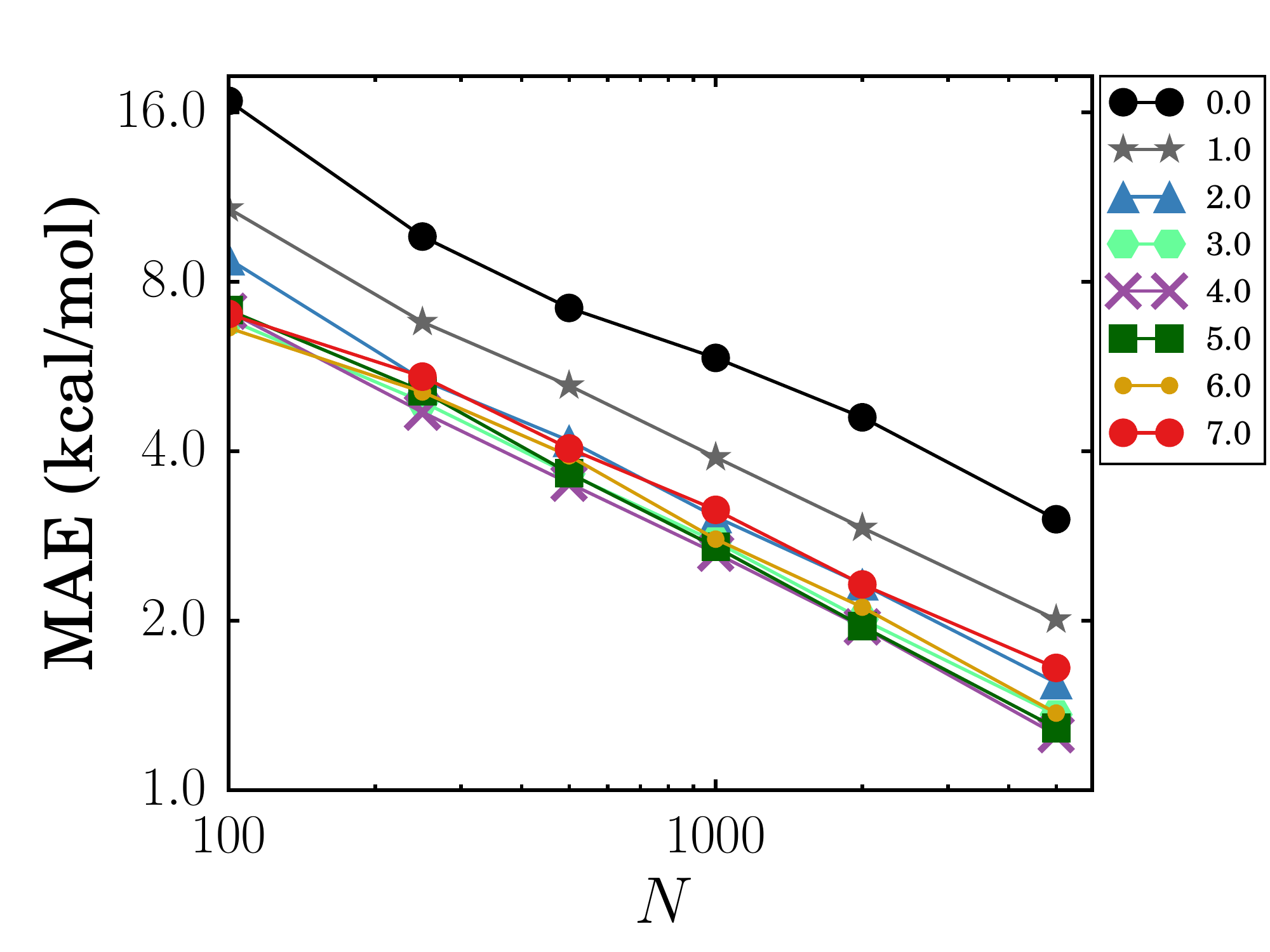}
 \includegraphics[width=0.45\linewidth]{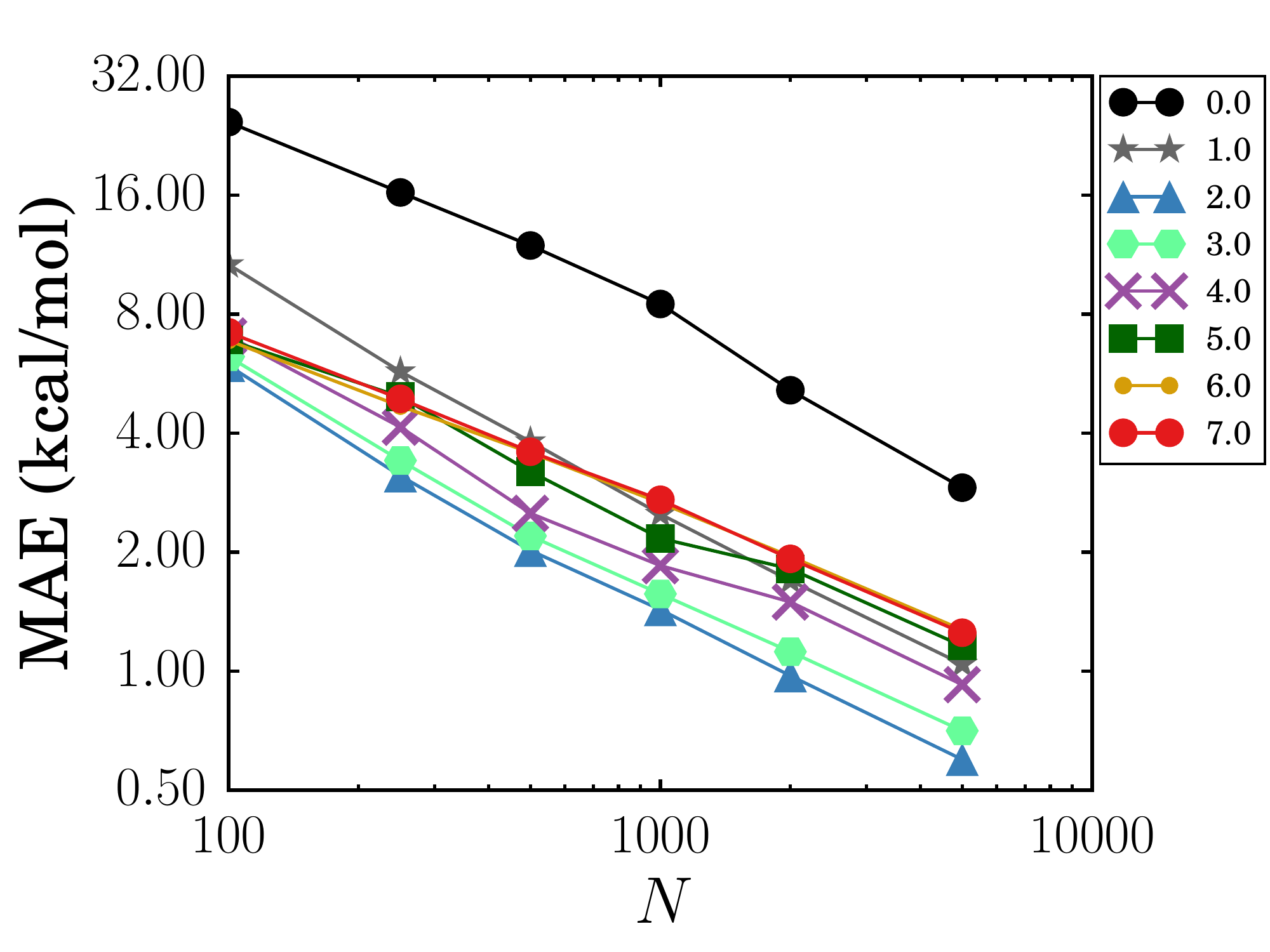}
 \caption{ \label{fig:2blc}
  Optimization of exponents in scaling power laws. 
    LEFT: Out-of-sample MAE for atomization/formation energy predictions
    as a function of training set size on the QM9 data set.
    Learning curves are generated using KRR with $\mathcal{A}_2$ as
    representation. 
    The legends indicate the exponent $n2$ used in the scaling power law, $\xi_2(d)$. 
     RIGHT: Out-of-sample MAE for atomization/formation energy predictions
    as a function of training set size on the QM9 data set.
    Learning curves are generated using KRR with $\mathcal{A}_3$ as
    representation. 
    The legends indicate the exponent $n3$ used in the scaling power law, $\xi_3(d)$. 
}
\end{figure}

We have screened screened radial exponents for the scaling functions $\xi_2(d_{iI})  = \dfrac{1}{d_{iI}^{n2}}$ and $\xi_3(d_{iI},d_{jI},\theta_{ij}^I) = \dfrac{1 - 3\cos(\theta_{ij}^I)\cos(\theta_{Ij}^i)\cos(\theta_{iI}^j)}{(d_{iI} d_{jI} d_{ij})^{n3}}$ , using atomization energies for  a subset of the QM9 dataset in order to identify
the optimal exponents.
Corresponding learning curves are shown in Fig.~\ref{fig:2blc}.
First, we have screened $\xi_2$, using $\mathcal{A}_2$ as representation, yielding the lowest
off-set for $n2=4$. Keeping this exponent for $\xi_2$ fixed, we then proceeded to screen the exponent $\xi_3$ in $\mathcal{A}_3$
We found that $n3 = 4$ corresponded to the best exponents for $\xi_3$.  We have used these values throughout this work, and unless something else is specified,
the optimal scaling functions read,
\begin{equation}
\begin{aligned}
\label{eq:scalingFunctions}
\xi_2(d_{iI}) & = \dfrac{1}{d_{iI}^4}\\
\xi_3(d_{iI},d_{jI},\theta_{ij}^I) &= \dfrac{1 - 3\cos(\theta_{ij}^I)\cos(\theta_{Ij}^i)\cos(\theta_{iI}^j)}{(d_{iI} d_{jI} d_{ij})^2}
\end{aligned}
\end{equation}

\subsubsection{Alchemical smearing}
\begin{figure}
 \includegraphics[width=0.95\linewidth]{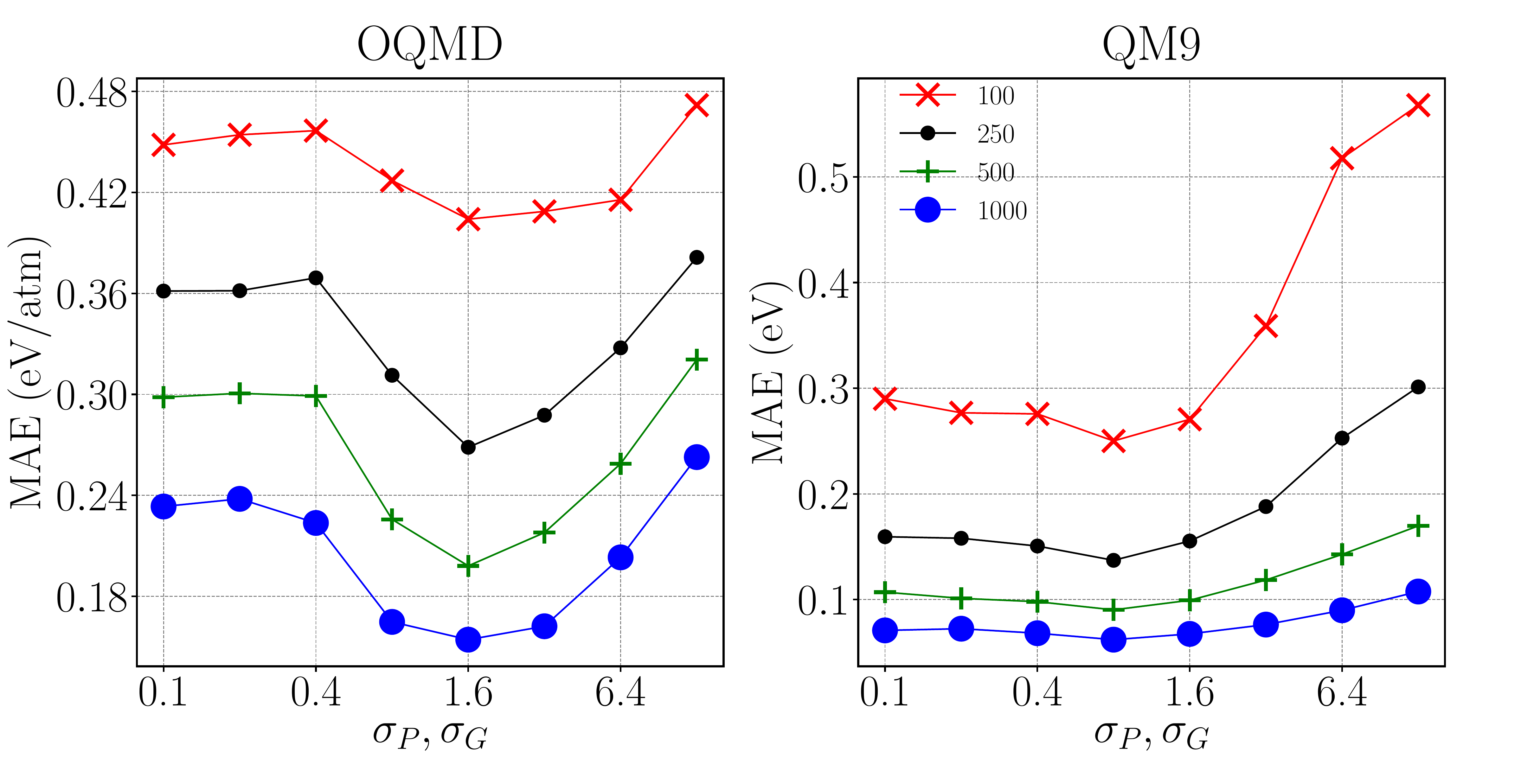}
 \caption{ \label{fig:AlchsSreen}
    Changes in out-of-sample MAE as a function of uniform Gaussian width ($\sigma_P$ and $\sigma_G$) used for elemental smearing.
    Results for energy predictions in the OQMD (LEFT) and QM9 (RIGHT) datasets, respectively. 
    Legends indicate the training set size.
}
\end{figure}

Parameters associated with the elemental smearing have also turned out to have 
a strong effect on the predictive power of the QML models.
We have therefore screened the corresponding values of $\sigma_P$ and $\sigma_G$ using 
energy prediction errors for the OQMD and QM9 data set for different training set sizes. 
These two datasets have been used due to their (relatively) high (OQMD) and low (QM9) chemical diversity in terms of
number of differing elements in the the data set. 
The optimal alchemical Gaussian widths varies only slightly across the two sets, 
as shown in Fig.~\ref{fig:AlchsSreen}.
A circular Gaussian with width $\sigma_P =\sigma_G = \sim$1.6,
which amounts to $\sim$90\% overlap between neighboring elements,
corresponds in a relatively deep well with minimal MAE for the OQMD dataset, no matter the training set size.
The fact that the optimal width stays constant with respect to training set size is beneficial:
The elemental smearing can be optimized using relatively small training sets, 
and can then be applied to larger training sets.
Comparing the MAE from a model with $\sigma_P = \sigma_G = 0.1$ (which in practice is equivalent zero overlap between different atomic types), using the optimal $\sigma_P = \sigma_G$
lowers the MAE by $\sim$9.9\% for the OQMD data set at 100 training samples, which increases up to $\sim$34\% when 1k training samples are used. 
Prediction errors for the QM9 data set indicate similar behavior, yet much less pronounced. 
For the largest  training set (1000 molecules), the optimization well becomes very shallow, 
consistent with the lack of compositional diversity in QM9.

Unsurprisingly, datasets with higher chemical diversity benefit more from using
the optimized elemental widths.  It may therefore not always be beneficial to include any elemental overlap,
especially for datasets with low elemental diversity, as it is computationally more expensive to do so.

\section{Data sets}
\label{sec:data}
We have used multiple datasets to benchmark out-of-sample accuracy of energy predictions of our model. 
These datasets includes organic molecules, crystals, biomolecular dimers, water clusters, and main-group diatomics. 
Some of the datasets are high-quality, have already been published and are in widespread use.
Additional low quality data sets have been generated, merely in order to accumulate 
additional evidence for the relative improvement of the new representation.
Since test set predictions are always close to zero by construction,
we exclusively report prediction errors as out-of-sample errors 
(averaged through cross-validation) with respect to reference validation numbers. 
All errors reported correspond to at least 10 cross-validation runs for each training set size. 

\subsection{Organic molecules: QM9}
The QM9 dataset~\cite{gdb9} corresponds to hybrid DFT~\cite{B3LYP} based
structures and properties of 134k organic molecules with up to 
nine atoms (C, O, N, or F), not counting hydrogen. 
SMILES strings of these molecules correspond to 
a subset of the GDB-17 dataset~\cite{gdb17}. 
The 3k organic molecules, which fail SMILES consistency tests~\cite{gdb9}, 
were removed before use.

A random subset of 22k molecules was selected from QM9 for training and testing. 
2k molecules were used for testing, and up to 20k for training.

\subsection{Organic molecules: QM7b}
Due to widespread use we also included the more established QM7b dataset~\cite{Montavon2013MLchemicalCompSpace}. 
QM7b was also derived from GDB~\cite{gdb13}.  
It contains hybrid DFT (PBE0~\cite{PBE01,PBE02}) structures and properties of $\sim$7k organic molecules with up to seven atoms
(C, O, N, S or Cl), not counting H.
We have drawn at random up to 5k molecules for training, and 2k for testing.

\subsection{Biomolecular dimers: SSI}
For intra-molecular and non-equilibrium interactions we used a subset
of 2356 neutral dimers from recently published protein-sidechain-sidechain interaction (SSI) dataset~\citet{BurnsSSI}. 
The SSI dataset is a collection of dimers mimicking configurations of interacting amino-acid sidechains as observed in a set of 47 high-resolution protein 
crystal structures.  The energies correspond to the DW-CCSD(T**)-F12 level of theory~\cite{Marshall2011}.

\subsection{Water cluster}
We also include a dataset which we calculated for 4'000 snapshots drawn from a molecular dynamics trajectory of a water cluster consisting of 40 water molecules. 
For the molecular dynamics we used the NET-ensemble at 300K, the Tip3p potential~\cite{Tip3p}, as implemented in CHARMM C41a1~\cite{brooks1983charmm}.
The energies correspond to an optimized combination of basis-sets and DFT functional (PBEh-3c)~\cite{grimme2015consistent}, 
as implemented in Orca~\cite{neese2012orca}, obtained for each of the 4'000 geometries.
%More details can be found in the SI.
%We compare the results to CM, BOB SLATM and aSLATM with KRR.

\subsection{Solids: OQMD}
We have used the Inorganic Crystal Structure Database~\cite{icsd,icsd2} subset
corresponding to the open quantum materials data base (OQMD) by Wolverton and co-workers~\cite{kirklin2015open,Saal2013}.
This data-set has already been used to develop and benchmark random forest based QML model (Voronoi)~\citet{Ward2017Voronoi}.
The dataset consists of $\sim$30k crystal structures and formation energies, calculated using high-throughput DFT (GGA+U).
We have used a random subset consisting of 3k structures with less 
than 40 atoms in the unit cell and formation energies lower than $5$ eV/atom 
for training and testing.
1k crystals were used for testing, and up to 2k for training. 

\subsection{Solids: Elpasolites}
We have also tested our representation for the 
Elpasolite crystal structure data set~\cite{faber2016}. 
This data set consists of $\sim10$k Elpasolite structures and DFT (PBE~\cite{PBE}) formation energies. 
The crystals correspond to quaternary main group elemental composition with all elements up to Bismuth (39 in total).
We have used a random subset of 7k structures, with up to 6k and 1k for training and testing, respectively. 

\subsection{Maingroup diatomics}
To test the predictive power for alchemical interpolation
we have also included a set of previously published DFT (PBE~\cite{PBE}) results for
single, double, and triple bonds among main-group diatomics
saturated with hydrogens~\cite{Samuel2016}.

\section{Results and Discussion}
\label{sec:results}
Using learning curves (always resulting in straight lines when recorded on log-log plots due to their inverse power law relationship~\cite{StatError_Muller1996}), 
we first present numerical results which indicate the 
predictive power of our QML model for atomization and formation energies in various data sets. 
When available for the same data set, we also compare to other QML models in the literature. 
Thereafter, the alchemical extrapolation capacity is 
demonstrated for predicting covalent bonds in molecules 
with elements that were not part of training. 
Finally, log-log plots of learning curves for nine electronic ground-state properties of organic
molecules (QM9) are reported and discussed.

\subsection{Energies of molecules, clusters, and solids}

\begin{figure}
 \includegraphics[width=0.95\linewidth]{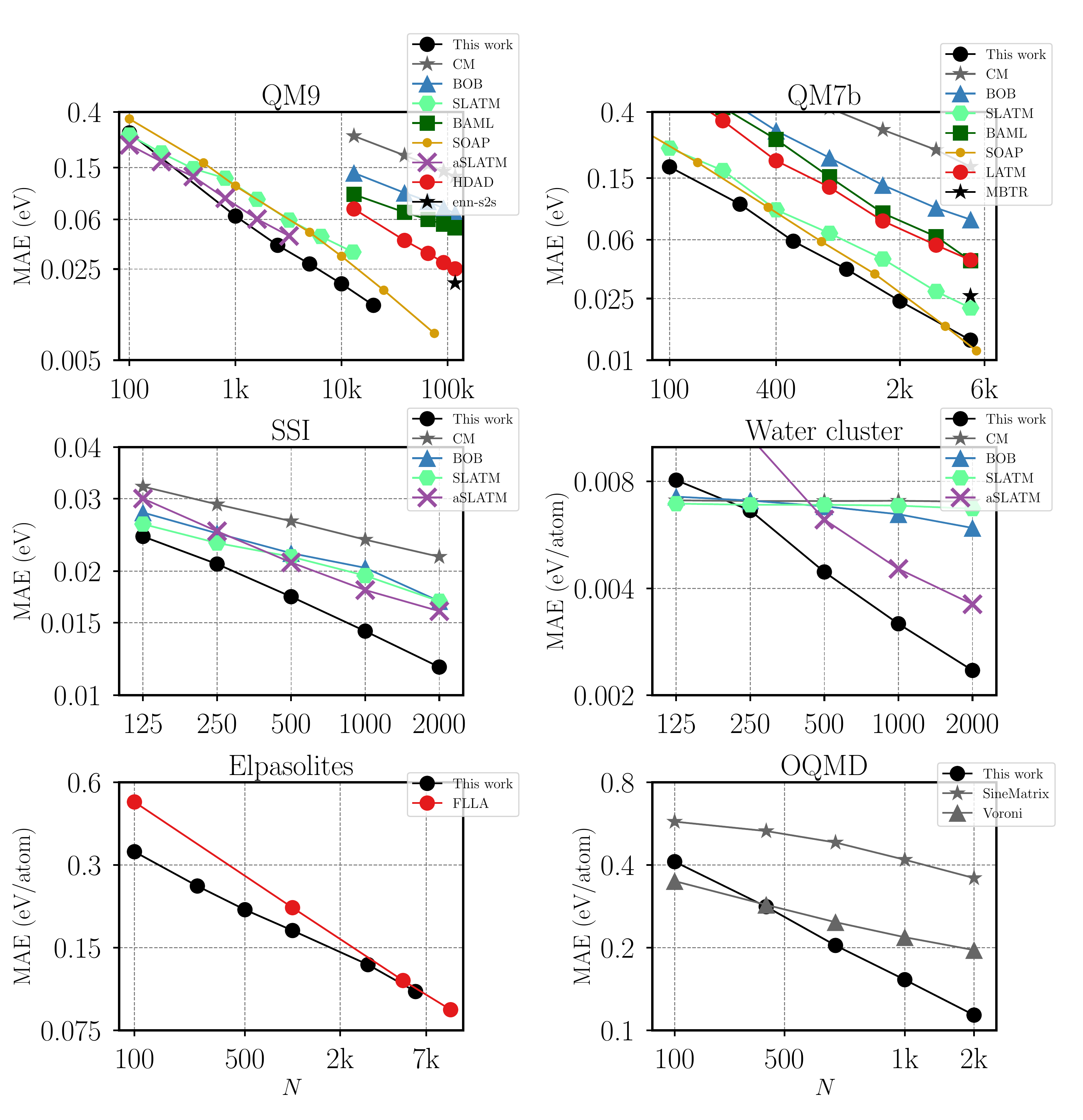}
 \caption{ \label{fig:LearningCurve}
  Learning curves for atomization/formation energy predictions corresponding to various QML models.
  Out-of-sample MAE is shown as a function of training set size for molecular (QM9 and QM7b), protein side-chain dimers (SSI), 
    liquid water ((H$_2$O)$_{40}$ snapshots (Water cluster) 
    and crystalline (OQMD and Elpasolites) data-sets. 
     }
\end{figure}

Fig.~\ref{fig:LearningCurve} displays the performance overview for energy predictions 
on six different data sets (QM9, QM7b, SSI, water, elapsolites, OQMD). 
Mean absolute out-of-sample energy prediction errors are shown as a function of training set size. 
The results indicate remarkable performance for all data sets, indicating a well working
QML model yielding systematic improvement with increasing training set size. 
The learning curves also indicate out-of-sample MAEs which are consistently lower, or similar, than 
previously published models in the literature. 
For QM9, the MAE reaches the highly coveted chemical accuracy threshold 
($1$ kcal/mol or $\sim0.043$ eV for enthalpy of formation)
with only 2k training points on the QM9 dataset.
Previously published QML models had to include an order of magnitude more training molecules to reach such accuracy.
This is similar to the amount of training molecules necessary when using the Coulomb matrix representation in conjunction
with semi-empirical or DFT based baselines in order to estimate electron correlated energies, as demonstrated in 2015 with
the $\Delta$-ML model~\cite{deltalearning}. 

For QM9, aSLATM~\cite{amons} and SOAP multi kernel model~\cite{bartok2017machine,SOAP_apl} reach a performance 
nearly as good as our QML model. aSLATM, however, performs worse for the SSI and the Water cluster.
The SOAP multi kernel QML model, however, performs an expansion in kernel function space acting on the distance
for which all degrees of freedom have already been integrated out. As such it is, strictly speaking,
not the same as as an improved representation, but rather an improved regressor. 
Note that single kernel based SOAP QML models perform significantly worse.
The reader should take notice however that in the SOAP learning curve results presented in 
Fig~\ref{fig:LearningCurve}, the $\sim$3k structures which had failed the SMILES consistency test,
were included. As such, theses QML models are not exactly comparable, and the SOAP results are still
likely to slightly improve if these faulty structures were to be removed.
One should also note that the SOAP results shown for QM7b
correspond to the multi-kernel SOAP kernel~\cite{SOAP_original,SOAP_apl}.

Other models presented correspond to Coulomb matrix (CM)~\cite{ML0}, bags of bonds (BOB)~\cite{bob}, 
Bonds and Angles based Machine Learning (BAML)~\cite{BAML}, Histogram of Distances, Angles, and
Dihedrals (HDAD)~\cite{faber2017fast}, Spectral London Axilrod-Teller-Muto (SLATM), atomic SLATM (aSLATM)~\cite{amons},
the crystal representation by Faber, Lindmaa, Lilienfeld, Armiento (FLLA)~\cite{faber2016}, the Sinematrix~\cite{qc_Felix_crystal},
and the unified many-body tensor representation (MBTR)~\cite{huo2017unified}. 
We also compared to QML models which are not based on KRR, such as the
message passing neural network model (enn-s2s)~\cite{NeuralMessagePassing},
and a Voronoi-tessellation based random forest model (Voronoi)~\citet{Ward2017Voronoi}. 

The MAE of our new QML model is consistently the lowest for all data sets and large training sets. 
For the set of 4,000 non-equilibrium water clusters, there is a noticeable difference between the global (CM, BOB and SLATM)
and the atomic representations (i.e., aSLATM and the new model we introduce in this work):
The global models exhibit very little learning at first, only for larger $N$ the learning curves begin to turn downward.
The atomic models, however, our new representation based QML model as well as aSLATM, improve rapidly with increasing training data set size.
We believe that sorting and crowding in the global representations makes it difficult to accurately account for the purely 
geometrical changes in structures that contribute to total energy variations.

Impressive predictive power is also observed for the OQMD dataset, a structurally 
and compositionally very diverse set of solids.
Our new model has a lower out-of-sample MAE for all $N$ 
when compared to the sine matrix representation on the OQMD dataset.
The offset of the learning curve of our new model is larger 
compared to that of the Voroni-based random-forest model~\citet{Ward2017Voronoi}.
However, the learning rate of our QML model is significantly steeper, 
surpassing the Voronoi model already  at just $\sim 250$ training samples. 
Results for a solid state variant of the CM, designed for use in periodic systems,
has also been included (SineMatrix)~\cite{qc_Felix_crystal}.
It has a similar slope as the Voronoi model, but a substantially larger off-set.

For the elpasolite data set,~\cite{faber2016}, with large composition diversity
but identical crystal structures, the learning-curve of the FLLA representation 
has a slightly higher off-set than our new QML model, yet exhibits a steeper 
learning curve. Our model converges towards the same slope for larger training set sizes.
We can only speculate on the reasons for such behavior.
The FLLA representation differs qualitatively from the 
other representations in this study: It does not 
include any explicit information about coordinates
and only encodes periodic row and column of the
elements which populate each crystal structure site. 
The QML model then learns to infer ground state energies
without knowing the exact configuration.
This leads to a very low dimensional model that is 
still unique for the system, 
which might be the cause of the lower slope. 
This however needs to be investigated more carefully
before any conclusions can be drawn.

\subsection{Alchemical predictions}

\begin{figure}
 \includegraphics[width=0.95\linewidth]{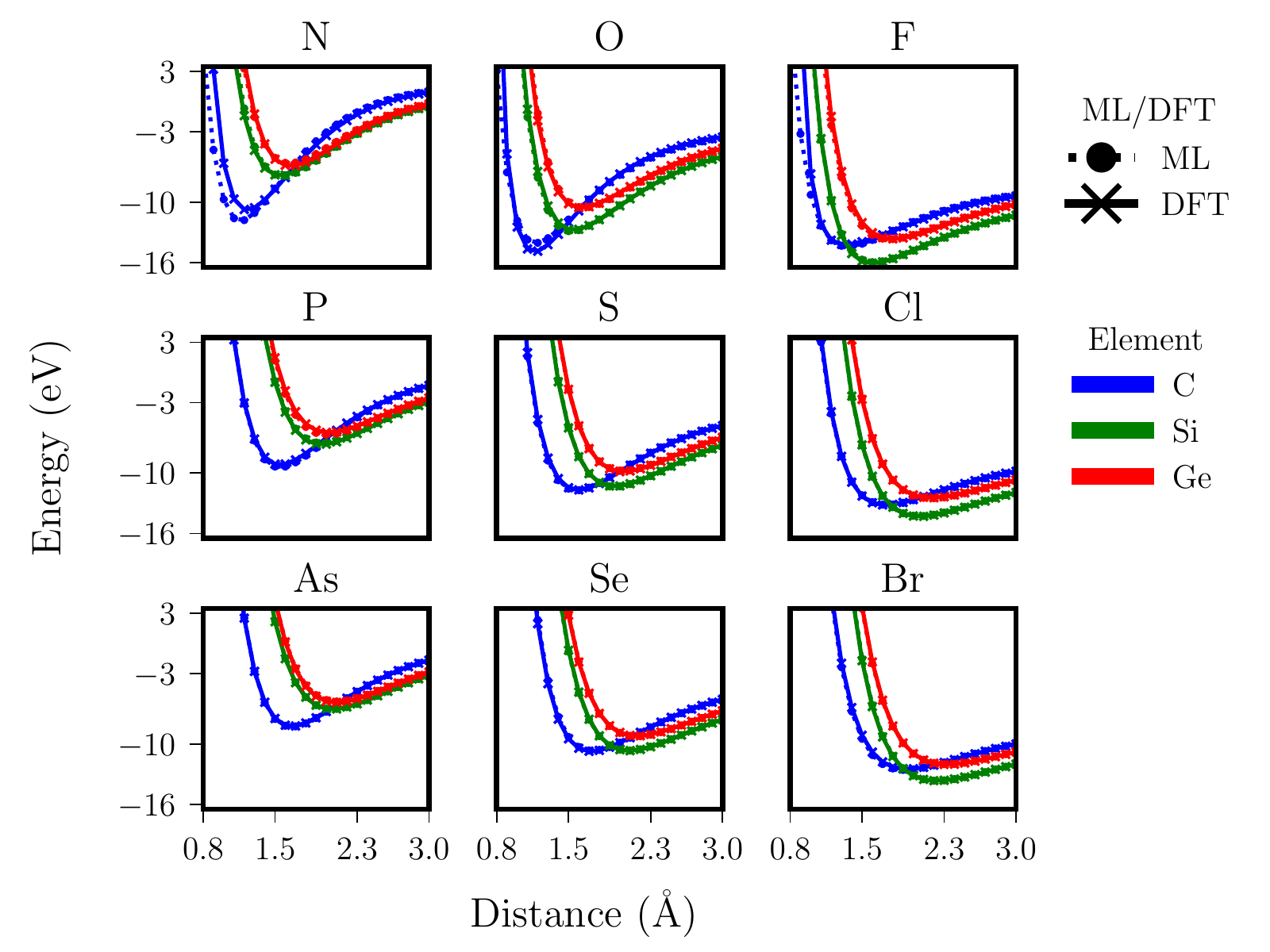}
 \caption{ \label{fig:alchExtrapolate}
    Covalent bond potentials calculated by DFT (star) and estimated by QML (circle) for 27 main-group diatomic molecules. 
    Bonding occurs between a group IV element (C blue, Si green, or Ge red), and halogens (single bond), chalcogen (double bond), or a group V element (triple bond).
    Columns correspond to triple (LEFT), double (MID), and single bonds (RIGHT).
    Rows correspond to the period of the group IV atom's binding partner:
    2nd period (TOP), 3rd period (MID), 4th period (BOTTOM). 
}
\end{figure}

Our new scaled many-body expansion explicitly accounts not only for distributions of interatomic distances and angles but 
also for elemental distributions in the periodic table.
We have therefore studied its capability to predict covalent binding of molecules containing chemical elements which were
not present in the molecules used for training.
More specifically, we have investigated single, double, and triple bonds with one bonding atom coming from group (IV), i.e. C, Si, or Ge. 
In order to increase covalent bond order, we have varied the valency of their bonding partner as follows:
For single bonds, group IV atoms are bound to halogens (group VII). 
For double bonds, group IV atoms atoms are bound to chalcogen atoms (group VI), and for triple bonds, 
group IV atoms are bound to group V atoms.
Dangling valencies of group IV atoms have been saturated with hydrogen.
Similar covalent bonding potentials have also recently been used in order to assess the predictive power of 
first and second order perturbation theory based alchemical predictions~\cite{Samuel2016}.

In order to test the alchemical ``extrapolation'', we trained on the covalent bonds of all other compounds (16 curves) 
which did contain neither the group IV atom nor the corresponding bonding partner in question.
The predictive power for the out-of-sample molecule, on display in Fig.~\ref{fig:alchExtrapolate}, is impressive. 
Albeit not quantitative (chemical accuracy is not reached), the results are semi-quantitative and certainly 
provide a physically very adequate picture of the covalent bonding in single, double, and triple bonds
for main-group atoms in periods 2 to 4.
The fact that predictions for the central elements H$_2$SiS are more accurate (easier to interpolate)
than others is consistent with this interpretation.
We also note that the deviation is the worst for 2nd-row elements (due to lack of $d$-orbitals
they differ substantially more from 3rd and 4th row than 3rd and 4th row differ from each other). 
Because of their poor performance we have not included other representations in this test.

These results clearly demonstrate that alchemical extrapolation is possible when interpolating
elemental groups and periods in the periodic table through an appropriate representation.
Since the representation is continuous in the corresponding compositional space, we also believe
that indication is given that the calculation of alchemical derivatives is meaningful, 
similar in spirit to Ref.~\cite{anatole-jcp2009-2}.

%We then conciser 
%contains $X Y$ bond potentials (atomization energies) of H$_{n-4} X Y$ molecules;
%where $X$ is C, Si or Ge;
%$Y$ is N, O, F, P, S, Cl, As, Se or Br;
%and $n$ is the number of valence electrons in $Y$.
%The bond potentials in the figure have been calculated using DFT,
%and predicted with QML.
%ML predictions on a given $X Y$ bond potential is trained
%on all other $X' Y'$ bond potentials in the dataset,
%and QML model is made using the same hyperparameters used for the QM9 data set (see SI for details).
%
%The predicted bond potentials agrees qualitatively with the DFT potentials,
%which demonstrates that our model can infer the behavior of an element 
%without needing to observe it.
%This is of monumental importance 
%when working with chemically diverse data sets, 
%where binned representations will inevitable fail due to 
%the combinatorial growth of bins with respect to the number 
%of elements in the data set.

\subsection{Other ground state properties of molecules}
\begin{figure}
 \includegraphics[width=0.95\linewidth]{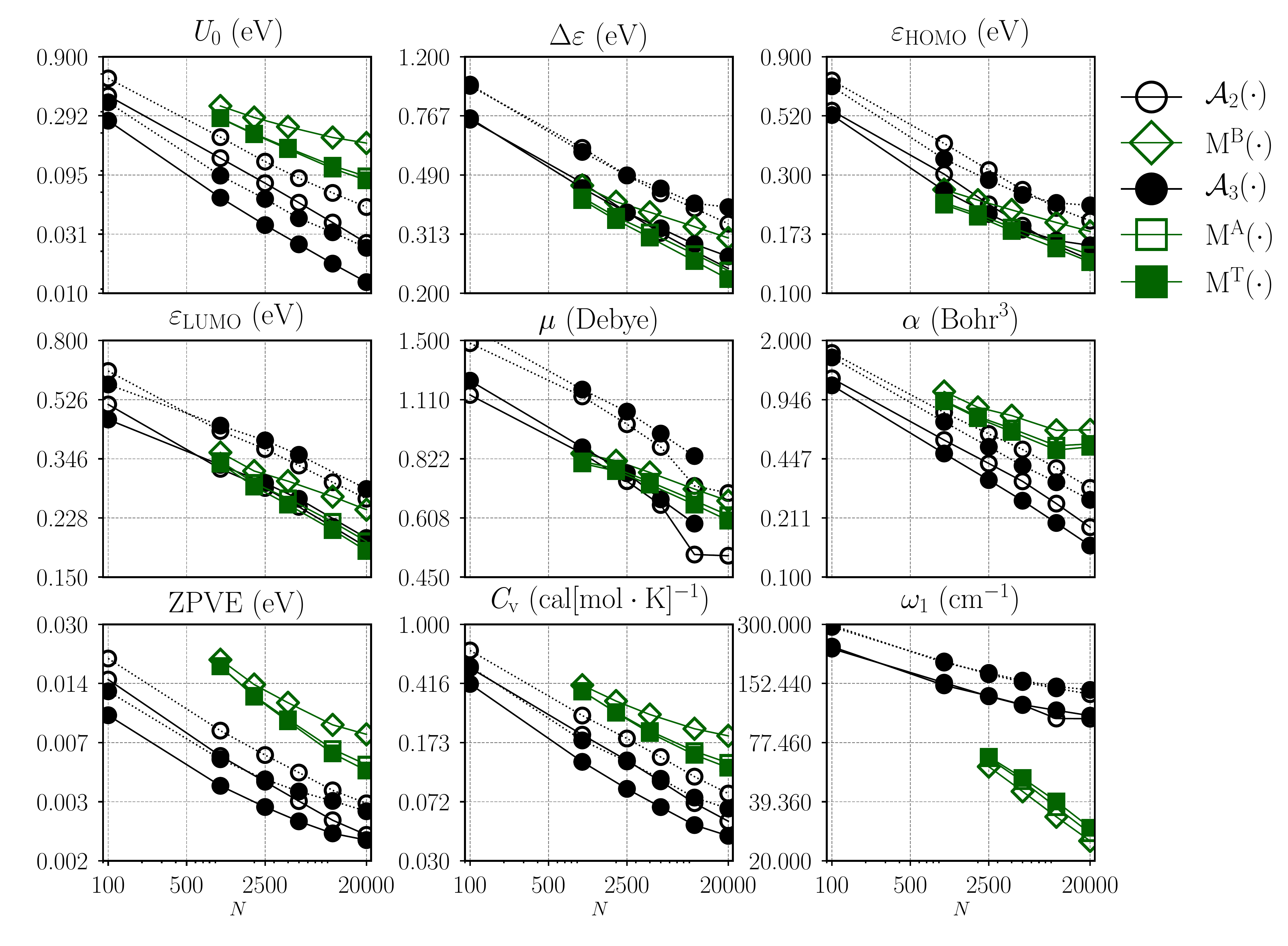}
 \caption{ \label{fig:qm9AllProps}
    Learning curves for out-of-sample MAE (filled lines) and RMSE (dashed lines) as a
    function of training set size $N$ for nine electronic ground state properties in the QM9 dataset.
    QML predictions have been made using either a molecular kernel
    and BAML as representation, or atomic kernels with our new representation.
    The BAML representation includes bonds ($M^B$); bonds and angles ($M^A$);
    and bonds, angles and torsional angles ($M^T$).
    Predicted properties include:
    atomization energy, at 0 Kelvin ($\mathit{U}_{\rm 0}$);
    HOMO-LUMO gap ($\Delta\varepsilon$);
    HOMO eigenvalue ($\varepsilon_{\mathrm{HOMO}}$);
    LUMO eigenvalue ($\varepsilon_{\mathrm{LUMO}}$);
    norm of dipole moment ($\mu$); 
    static isotropic polarizability ($\alpha$);
    zero point vibrational energy (ZPVE);
    heat capacity at room temperature ($\mathit{C}_{\mathrm{v}}$);
    and the highest fundamental vibrational frequency ($\omega_1$). 
    }
\end{figure}

Finally, we also investigated how well QML models based on our new representation, optimized for energies, 
performs for predicting other ground-state quantum properties part of the QM9 dataset.
More specifically, we have included atomization energies, HOMO, LUMO-eigenvalues as well as gap, 
dipole moment, polarizability, zero point vibrational energy, heat capacity, and the vibrational 
frequency of the highest lying fundamental ($\omega$).
Results are shown in Fig.~\ref{fig:qm9AllProps}, and provide overwhelming evidence 
that resulting models enable predictions systematically improving with training set size, no matter what property.
For comparison, we have also included results for the aSLATM model.
aSLATM results are typically worse when dealing with extensive properties, such as
energies, polarizability, or heat-capacity. 
When dealing with intensive properties, such as eigenvalues or dipolemoment, aSLATM is on par or even slightly better than our model, 
with the exception of $\omega$.
$\omega$ corresponds to frequency associated to the vibrational stretch of CH, NH, or OH bonds, 
a property with hardly any variance at all. 
Previously we have seen that this property is best predicted by a random forest model which 
has poor performance for all other properties~\cite{faber2017fast}.
The interpretation is that predicting this property is much more a 
classification problem, then a supervised learning task.

Fig.~\ref{fig:qm9AllProps} also includes learning curves for the root mean squared error, indicating
the slightly higher offset than the mean absolute error, to be expected, and systematic improvement with
training set size with similar slopes as the mean absolute error. 
This is an assuring result, indicating once again, that also predictions for outliers improve
as training set size is increased~\cite{deltalearning}.

Furthermore, for Fig.~\ref{fig:qm9AllProps} we have also distinguished 
between two and three body contributions (as well as four-body for BAML).
For all properties but for $\omega$ the trend meets the expectation, as also 
already confirmed previously for BAML~\cite{BAML}: 
Addition of the higher order term systematically lowers the learning curves by a significant amount.

\section*{Conclusion}
We have introduced a universal representation of an atom in a chemical compound for use in QML models.
An atom is represented by a sum of multidimensional Gaussians,
each term corresponding to elemental, atom-pairwise, and angular distributions and scaled by respective power laws.
For the compounds and properties studied we have found four-body contributions to be insignificant.
System-independent hyperparameters, such as exponents in scaling functions and Gaussian widths have been optimized
using the out-of-sample prediction error for the energy as a penalty.
Analytical expressions have been derived for corresponding distances between arbitrary chemical compounds.
These distances can directly be used within kernel ridge regression based QML models of electronic ground state properties.
For energies of organic molecules, water clusters, amino-acid side chains, and crystalline solids 
the resulting QML models lead to learning curves with very low off-set and steep learning rate.
For compositionally diverse systems chemical accuracy ($\sim$1 kcal/mol) can now be reached 
using only thousands of training instances.
We have also studied the effect of explicitly accounting for inter-elemental distances in the periodic table:
Our new QML model can produce semi-qualitatively accurate covalent bonding potentials for single, double, and triple 
bonds which include chemical element-pairs which were not part of training. 
For various electronic ground state properties of organic molecules, numerical results indicate that 
our new QML model has remarkable predictive power. 

While the reference data used in this study has mostly been obtained at the hybrid DFT level of theory, 
the steep learning curves of our QML models suggest that it has now become a realistic possibility 
to obtain a sufficiently large training set at post-Hartree-Fock level of theory (or from experiment), 
and to use it for the training of QML models which enable subsequent high-throughput screening efforts with similar accuracy.

Combining our new representation with the recently proposed amon approach will provide
the possibility to obtain highly accurate QML models which are scale invariant, i.e.~they 
can be applied to systems containing arbitrarily many atoms~\cite{amons}.
Future work will deal with forces and other properties.

\section*{Acknowledgement}
F.A.F. would like to thank Kuang-Yu Samuel Chang and Michele Ceriotti for data provided.
O.A.v.L. acknowledges funding from the Swiss National Science foundation (No.~PP00P2\_138932 and 407540\_167186 NFP 75 Big Data).
This research was partly supported by the NCCR MARVEL, funded by the Swiss National Science Foundation.
Calculations were performed at sciCORE (http://scicore.unibas.ch/) scientific computing core facility at University of Basel.

\bibliography{ref}

\begin{thebibliography}{55}%
\makeatletter
\providecommand \@ifxundefined [1]{%
 \@ifx{#1\undefined}
}%
\providecommand \@ifnum [1]{%
 \ifnum #1\expandafter \@firstoftwo
 \else \expandafter \@secondoftwo
 \fi
}%
\providecommand \@ifx [1]{%
 \ifx #1\expandafter \@firstoftwo
 \else \expandafter \@secondoftwo
 \fi
}%
\providecommand \natexlab [1]{#1}%
\providecommand \enquote  [1]{``#1''}%
\providecommand \bibnamefont  [1]{#1}%
\providecommand \bibfnamefont [1]{#1}%
\providecommand \citenamefont [1]{#1}%
\providecommand \href@noop [0]{\@secondoftwo}%
\providecommand \href [0]{\begingroup \@sanitize@url \@href}%
\providecommand \@href[1]{\@@startlink{#1}\@@href}%
\providecommand \@@href[1]{\endgroup#1\@@endlink}%
\providecommand \@sanitize@url [0]{\catcode `\\12\catcode `\$12\catcode
  `\&12\catcode `\#12\catcode `\^12\catcode `\_12\catcode `\%12\relax}%
\providecommand \@@startlink[1]{}%
\providecommand \@@endlink[0]{}%
\providecommand \url  [0]{\begingroup\@sanitize@url \@url }%
\providecommand \@url [1]{\endgroup\@href {#1}{\urlprefix }}%
\providecommand \urlprefix  [0]{URL }%
\providecommand \Eprint [0]{\href }%
\providecommand \doibase [0]{http://dx.doi.org/}%
\providecommand \selectlanguage [0]{\@gobble}%
\providecommand \bibinfo  [0]{\@secondoftwo}%
\providecommand \bibfield  [0]{\@secondoftwo}%
\providecommand \translation [1]{[#1]}%
\providecommand \BibitemOpen [0]{}%
\providecommand \bibitemStop [0]{}%
\providecommand \bibitemNoStop [0]{.\EOS\space}%
\providecommand \EOS [0]{\spacefactor3000\relax}%
\providecommand \BibitemShut  [1]{\csname bibitem#1\endcsname}%
\let\auto@bib@innerbib\@empty
%</preamble>
\bibitem [{\citenamefont {Jensen}(2007)}]{JensenCompChem}%
  \BibitemOpen
  \bibfield  {author} {\bibinfo {author} {\bibfnamefont {F.}~\bibnamefont
  {Jensen}},\ }\href@noop {} {\emph {\bibinfo {title} {Introduction to
  Computational Chemistry}}}\ (\bibinfo  {publisher} {John Wiley},\ \bibinfo
  {address} {West Sussex, England},\ \bibinfo {year} {2007})\BibitemShut
  {NoStop}%
\bibitem [{\citenamefont {Brockherde}\ \emph {et~al.}(2017)\citenamefont
  {Brockherde}, \citenamefont {Vogt}, \citenamefont {Li}, \citenamefont
  {Tuckerman}, \citenamefont {Burke},\ and\ \citenamefont
  {M{\"u}ller}}]{brockherde2017bypassing}%
  \BibitemOpen
  \bibfield  {author} {\bibinfo {author} {\bibfnamefont {F.}~\bibnamefont
  {Brockherde}}, \bibinfo {author} {\bibfnamefont {L.}~\bibnamefont {Vogt}},
  \bibinfo {author} {\bibfnamefont {L.}~\bibnamefont {Li}}, \bibinfo {author}
  {\bibfnamefont {M.~E.}\ \bibnamefont {Tuckerman}}, \bibinfo {author}
  {\bibfnamefont {K.}~\bibnamefont {Burke}}, \ and\ \bibinfo {author}
  {\bibfnamefont {K.-R.}\ \bibnamefont {M{\"u}ller}},\ }\href@noop {}
  {\bibfield  {journal} {\bibinfo  {journal} {Nature Communications}\ }\textbf
  {\bibinfo {volume} {8}},\ \bibinfo {pages} {872} (\bibinfo {year}
  {2017})}\BibitemShut {NoStop}%
\bibitem [{\citenamefont {Faber}\ \emph {et~al.}(0)\citenamefont {Faber},
  \citenamefont {Hutchison}, \citenamefont {Huang}, \citenamefont {Gilmer},
  \citenamefont {Schoenholz}, \citenamefont {Dahl}, \citenamefont {Vinyals},
  \citenamefont {Kearnes}, \citenamefont {Riley},\ and\ \citenamefont {von
  Lilienfeld}}]{faber2017fast}%
  \BibitemOpen
  \bibfield  {author} {\bibinfo {author} {\bibfnamefont {F.~A.}\ \bibnamefont
  {Faber}}, \bibinfo {author} {\bibfnamefont {L.}~\bibnamefont {Hutchison}},
  \bibinfo {author} {\bibfnamefont {B.}~\bibnamefont {Huang}}, \bibinfo
  {author} {\bibfnamefont {J.}~\bibnamefont {Gilmer}}, \bibinfo {author}
  {\bibfnamefont {S.~S.}\ \bibnamefont {Schoenholz}}, \bibinfo {author}
  {\bibfnamefont {G.~E.}\ \bibnamefont {Dahl}}, \bibinfo {author}
  {\bibfnamefont {O.}~\bibnamefont {Vinyals}}, \bibinfo {author} {\bibfnamefont
  {S.}~\bibnamefont {Kearnes}}, \bibinfo {author} {\bibfnamefont {P.~F.}\
  \bibnamefont {Riley}}, \ and\ \bibinfo {author} {\bibfnamefont {O.~A.}\
  \bibnamefont {von Lilienfeld}},\ }\href {\doibase 10.1021/acs.jctc.7b00577}
  {\bibfield  {journal} {\bibinfo  {journal} {Journal of Chemical Theory and
  Computation}\ }\textbf {\bibinfo {volume} {0}},\ \bibinfo {pages} {null}
  (\bibinfo {year} {0})},\ \bibinfo {note} {pMID: 28926232},\ \Eprint
  {http://arxiv.org/abs/http://dx.doi.org/10.1021/acs.jctc.7b00577}
  {http://dx.doi.org/10.1021/acs.jctc.7b00577} \BibitemShut {NoStop}%
\bibitem [{\citenamefont {Sch{\"u}tt}\ \emph {et~al.}(2017)\citenamefont
  {Sch{\"u}tt}, \citenamefont {Arbabzadah}, \citenamefont {Chmiela},
  \citenamefont {M{\"u}ller},\ and\ \citenamefont
  {Tkatchenko}}]{DeepTensorNN_2017}%
  \BibitemOpen
  \bibfield  {author} {\bibinfo {author} {\bibfnamefont {K.~T.}\ \bibnamefont
  {Sch{\"u}tt}}, \bibinfo {author} {\bibfnamefont {F.}~\bibnamefont
  {Arbabzadah}}, \bibinfo {author} {\bibfnamefont {S.}~\bibnamefont {Chmiela}},
  \bibinfo {author} {\bibfnamefont {K.~R.}\ \bibnamefont {M{\"u}ller}}, \ and\
  \bibinfo {author} {\bibfnamefont {A.}~\bibnamefont {Tkatchenko}},\
  }\href@noop {} {\bibfield  {journal} {\bibinfo  {journal} {Nat. Commun.}\
  }\textbf {\bibinfo {volume} {8}},\ \bibinfo {pages} {13890} (\bibinfo {year}
  {2017})}\BibitemShut {NoStop}%
\bibitem [{\citenamefont {Gilmer}\ \emph {et~al.}(2017)\citenamefont {Gilmer},
  \citenamefont {Schoenholz}, \citenamefont {Riley}, \citenamefont {Vinyals},\
  and\ \citenamefont {Dahl}}]{NeuralMessagePassing}%
  \BibitemOpen
  \bibfield  {author} {\bibinfo {author} {\bibfnamefont {J.}~\bibnamefont
  {Gilmer}}, \bibinfo {author} {\bibfnamefont {S.~S.}\ \bibnamefont
  {Schoenholz}}, \bibinfo {author} {\bibfnamefont {P.~F.}\ \bibnamefont
  {Riley}}, \bibinfo {author} {\bibfnamefont {O.}~\bibnamefont {Vinyals}}, \
  and\ \bibinfo {author} {\bibfnamefont {G.~E.}\ \bibnamefont {Dahl}},\ }in\
  \href@noop {} {\emph {\bibinfo {booktitle} {Proceedings of the 34nd
  International Conference on Machine Learning, {ICML} 2017}}}\ (\bibinfo
  {year} {2017})\BibitemShut {NoStop}%
\bibitem [{\citenamefont {Duvenaud}\ \emph {et~al.}(2015)\citenamefont
  {Duvenaud}, \citenamefont {Maclaurin}, \citenamefont {Iparraguirre},
  \citenamefont {Bombarell}, \citenamefont {Hirzel}, \citenamefont
  {Aspuru-Guzik},\ and\ \citenamefont {Adams}}]{duvenaud2015convolutional}%
  \BibitemOpen
  \bibfield  {author} {\bibinfo {author} {\bibfnamefont {D.~K.}\ \bibnamefont
  {Duvenaud}}, \bibinfo {author} {\bibfnamefont {D.}~\bibnamefont {Maclaurin}},
  \bibinfo {author} {\bibfnamefont {J.}~\bibnamefont {Iparraguirre}}, \bibinfo
  {author} {\bibfnamefont {R.}~\bibnamefont {Bombarell}}, \bibinfo {author}
  {\bibfnamefont {T.}~\bibnamefont {Hirzel}}, \bibinfo {author} {\bibfnamefont
  {A.}~\bibnamefont {Aspuru-Guzik}}, \ and\ \bibinfo {author} {\bibfnamefont
  {R.~P.}\ \bibnamefont {Adams}},\ }in\ \href@noop {} {\emph {\bibinfo
  {booktitle} {Advances in Neural Information Processing Systems}}}\ (\bibinfo
  {year} {2015})\ pp.\ \bibinfo {pages} {2215--2223}\BibitemShut {NoStop}%
\bibitem [{\citenamefont {Rupp}\ \emph {et~al.}(2012)\citenamefont {Rupp},
  \citenamefont {Tkatchenko},\ and\ \citenamefont {von Lilienfeld}}]{ML0}%
  \BibitemOpen
  \bibfield  {author} {\bibinfo {author} {\bibfnamefont {M.}~\bibnamefont
  {Rupp}}, \bibinfo {author} {\bibfnamefont {K.-R.}\ \bibnamefont {Tkatchenko},
  \bibfnamefont {Alexandre haand~M{\"u}ller}}, \ and\ \bibinfo {author}
  {\bibfnamefont {O.~A.}\ \bibnamefont {von Lilienfeld}},\ }\href
  {http://dx.doi.org/10.1103/PhysRevLett.108.058301} {\bibfield  {journal}
  {\bibinfo  {journal} {Phys. Rev. Lett.}\ }\textbf {\bibinfo {volume} {108}},\
  \bibinfo {pages} {058301} (\bibinfo {year} {2012})}\BibitemShut {NoStop}%
\bibitem [{\citenamefont {Hansen}\ \emph {et~al.}(2015)\citenamefont {Hansen},
  \citenamefont {Biegler}, \citenamefont {Ramakrishnan}, \citenamefont
  {Pronobis}, \citenamefont {von Lilienfeld}, \citenamefont {M{\"u}ller},\ and\
  \citenamefont {Tkatchenko}}]{bob}%
  \BibitemOpen
  \bibfield  {author} {\bibinfo {author} {\bibfnamefont {K.}~\bibnamefont
  {Hansen}}, \bibinfo {author} {\bibfnamefont {F.}~\bibnamefont {Biegler}},
  \bibinfo {author} {\bibfnamefont {R.}~\bibnamefont {Ramakrishnan}}, \bibinfo
  {author} {\bibfnamefont {W.}~\bibnamefont {Pronobis}}, \bibinfo {author}
  {\bibfnamefont {O.~A.}\ \bibnamefont {von Lilienfeld}}, \bibinfo {author}
  {\bibfnamefont {K.-R.}\ \bibnamefont {M{\"u}ller}}, \ and\ \bibinfo {author}
  {\bibfnamefont {A.}~\bibnamefont {Tkatchenko}},\ }\href {\doibase
  10.1021/acs.jpclett.5b00831} {\bibfield  {journal} {\bibinfo  {journal} {J.
  Phys. Chem. Lett.}\ }\textbf {\bibinfo {volume} {6}},\ \bibinfo {pages}
  {2326} (\bibinfo {year} {2015})}\BibitemShut {NoStop}%
\bibitem [{\citenamefont {Huang}\ and\ \citenamefont {von
  Lilienfeld}(2016)}]{BAML}%
  \BibitemOpen
  \bibfield  {author} {\bibinfo {author} {\bibfnamefont {B.}~\bibnamefont
  {Huang}}\ and\ \bibinfo {author} {\bibfnamefont {O.~A.}\ \bibnamefont {von
  Lilienfeld}},\ }\href {\doibase 10.1063/1.4964627} {\bibfield  {journal}
  {\bibinfo  {journal} {J. Chem. Phys}\ }\textbf {\bibinfo {volume} {145}},\
  \bibinfo {eid} {161102} (\bibinfo {year} {2016})}\BibitemShut {NoStop}%
\bibitem [{\citenamefont {Behler}\ and\ \citenamefont
  {Parrinello}(2007)}]{Behler2007Symmetry}%
  \BibitemOpen
  \bibfield  {author} {\bibinfo {author} {\bibfnamefont {J.}~\bibnamefont
  {Behler}}\ and\ \bibinfo {author} {\bibfnamefont {M.}~\bibnamefont
  {Parrinello}},\ }\href {\doibase 10.1103/PhysRevLett.98.146401} {\bibfield
  {journal} {\bibinfo  {journal} {Phys. Rev. Lett.}\ }\textbf {\bibinfo
  {volume} {98}},\ \bibinfo {pages} {146401} (\bibinfo {year}
  {2007})}\BibitemShut {NoStop}%
\bibitem [{\citenamefont {Rogers}\ and\ \citenamefont
  {Hahn}(2010)}]{rogers2010extended}%
  \BibitemOpen
  \bibfield  {author} {\bibinfo {author} {\bibfnamefont {D.}~\bibnamefont
  {Rogers}}\ and\ \bibinfo {author} {\bibfnamefont {M.}~\bibnamefont {Hahn}},\
  }\href@noop {} {\bibfield  {journal} {\bibinfo  {journal} {J. Chem. Inf.
  Model.}\ }\textbf {\bibinfo {volume} {50}},\ \bibinfo {pages} {742} (\bibinfo
  {year} {2010})}\BibitemShut {NoStop}%
\bibitem [{\citenamefont {von Lilienfeld}(2013)}]{anatole_qc2013}%
  \BibitemOpen
  \bibfield  {author} {\bibinfo {author} {\bibfnamefont {O.~A.}\ \bibnamefont
  {von Lilienfeld}},\ }\href {\doibase 10.1002/qua.24375} {\bibfield  {journal}
  {\bibinfo  {journal} {Int. J. Quantum}\ }\textbf {\bibinfo {volume} {113}},\
  \bibinfo {pages} {1676} (\bibinfo {year} {2013})}\BibitemShut {NoStop}%
\bibitem [{\citenamefont {Sch{\"u}tt}\ \emph {et~al.}(2014)\citenamefont
  {Sch{\"u}tt}, \citenamefont {Glawe}, \citenamefont {Brockherde},
  \citenamefont {Sanna}, \citenamefont {M{\"u}ller},\ and\ \citenamefont
  {Gross}}]{Muller_Gross_crystal}%
  \BibitemOpen
  \bibfield  {author} {\bibinfo {author} {\bibfnamefont {K.~T.}\ \bibnamefont
  {Sch{\"u}tt}}, \bibinfo {author} {\bibfnamefont {H.}~\bibnamefont {Glawe}},
  \bibinfo {author} {\bibfnamefont {F.}~\bibnamefont {Brockherde}}, \bibinfo
  {author} {\bibfnamefont {A.}~\bibnamefont {Sanna}}, \bibinfo {author}
  {\bibfnamefont {K.~R.}\ \bibnamefont {M{\"u}ller}}, \ and\ \bibinfo {author}
  {\bibfnamefont {E.~K.~U.}\ \bibnamefont {Gross}},\ }\href {\doibase
  10.1103/PhysRevB.89.205118} {\bibfield  {journal} {\bibinfo  {journal} {Phys.
  Rev. B}\ }\textbf {\bibinfo {volume} {89}},\ \bibinfo {pages} {205118}
  (\bibinfo {year} {2014})}\BibitemShut {NoStop}%
\bibitem [{\citenamefont {Huang}\ and\ \citenamefont {von
  Lilienfeld}(2017)}]{amons}%
  \BibitemOpen
  \bibfield  {author} {\bibinfo {author} {\bibfnamefont {B.}~\bibnamefont
  {Huang}}\ and\ \bibinfo {author} {\bibfnamefont {O.~A.}\ \bibnamefont {von
  Lilienfeld}},\ }\href@noop {} {\bibfield  {journal} {\bibinfo  {journal}
  {arXiv preprint arXiv:1707.04146}\ } (\bibinfo {year} {2017})}\BibitemShut
  {NoStop}%
\bibitem [{\citenamefont {Huo}\ and\ \citenamefont
  {Rupp}(2017)}]{huo2017unified}%
  \BibitemOpen
  \bibfield  {author} {\bibinfo {author} {\bibfnamefont {H.}~\bibnamefont
  {Huo}}\ and\ \bibinfo {author} {\bibfnamefont {M.}~\bibnamefont {Rupp}},\
  }\href@noop {} {\bibfield  {journal} {\bibinfo  {journal} {arXiv preprint
  arXiv:1704.06439}\ } (\bibinfo {year} {2017})}\BibitemShut {NoStop}%
\bibitem [{\citenamefont {von Lilienfeld}(2014)}]{TowardsCompoundDesign2014}%
  \BibitemOpen
  \bibfield  {author} {\bibinfo {author} {\bibfnamefont {O.~A.}\ \bibnamefont
  {von Lilienfeld}},\ }in\ \href@noop {} {\emph {\bibinfo {booktitle}
  {Many-Electron Approaches in Physics, Chemistry and Mathematics}}}\ (\bibinfo
   {publisher} {Springer},\ \bibinfo {year} {2014})\ pp.\ \bibinfo {pages}
  {169--189}\BibitemShut {NoStop}%
\bibitem [{\citenamefont {Chang}\ \emph {et~al.}(2016)\citenamefont {Chang},
  \citenamefont {Fias}, \citenamefont {Ramakrishnan},\ and\ \citenamefont {von
  Lilienfeld}}]{Samuel2016}%
  \BibitemOpen
  \bibfield  {author} {\bibinfo {author} {\bibfnamefont {K.~S.}\ \bibnamefont
  {Chang}}, \bibinfo {author} {\bibfnamefont {S.}~\bibnamefont {Fias}},
  \bibinfo {author} {\bibfnamefont {R.}~\bibnamefont {Ramakrishnan}}, \ and\
  \bibinfo {author} {\bibfnamefont {O.~A.}\ \bibnamefont {von Lilienfeld}},\
  }\href@noop {} {\bibfield  {journal} {\bibinfo  {journal} {The Journal of
  chemical physics}\ }\textbf {\bibinfo {volume} {144}},\ \bibinfo {pages}
  {174110} (\bibinfo {year} {2016})}\BibitemShut {NoStop}%
\bibitem [{\citenamefont {von Lilienfeld}\ \emph {et~al.}(2015)\citenamefont
  {von Lilienfeld}, \citenamefont {Ramakrishnan}, \citenamefont {Rupp},\ and\
  \citenamefont {Knoll}}]{OAvL_FRD}%
  \BibitemOpen
  \bibfield  {author} {\bibinfo {author} {\bibfnamefont {O.~A.}\ \bibnamefont
  {von Lilienfeld}}, \bibinfo {author} {\bibfnamefont {R.}~\bibnamefont
  {Ramakrishnan}}, \bibinfo {author} {\bibfnamefont {M.}~\bibnamefont {Rupp}},
  \ and\ \bibinfo {author} {\bibfnamefont {A.}~\bibnamefont {Knoll}},\ }\href
  {\doibase 10.1002/qua.24912} {\bibfield  {journal} {\bibinfo  {journal} {Int.
  J. Quantum}\ }\textbf {\bibinfo {volume} {115}},\ \bibinfo {pages} {1084}
  (\bibinfo {year} {2015})}\BibitemShut {NoStop}%
\bibitem [{\citenamefont {Faber}\ \emph {et~al.}(2016)\citenamefont {Faber},
  \citenamefont {Lindmaa}, \citenamefont {von Lilienfeld},\ and\ \citenamefont
  {Armiento}}]{faber2016}%
  \BibitemOpen
  \bibfield  {author} {\bibinfo {author} {\bibfnamefont {F.~A.}\ \bibnamefont
  {Faber}}, \bibinfo {author} {\bibfnamefont {A.}~\bibnamefont {Lindmaa}},
  \bibinfo {author} {\bibfnamefont {O.~A.}\ \bibnamefont {von Lilienfeld}}, \
  and\ \bibinfo {author} {\bibfnamefont {R.}~\bibnamefont {Armiento}},\ }\href
  {\doibase 10.1103/PhysRevLett.117.135502} {\bibfield  {journal} {\bibinfo
  {journal} {Phys. Rev. Lett.}\ }\textbf {\bibinfo {volume} {117}},\ \bibinfo
  {pages} {135502} (\bibinfo {year} {2016})}\BibitemShut {NoStop}%
\bibitem [{\citenamefont {De}\ \emph {et~al.}(2016)\citenamefont {De},
  \citenamefont {Bart\'ok}, \citenamefont {Cs\'anyi},\ and\ \citenamefont
  {Ceriotti}}]{SOAP_apl}%
  \BibitemOpen
  \bibfield  {author} {\bibinfo {author} {\bibfnamefont {S.}~\bibnamefont
  {De}}, \bibinfo {author} {\bibfnamefont {A.~P.}\ \bibnamefont {Bart\'ok}},
  \bibinfo {author} {\bibfnamefont {G.}~\bibnamefont {Cs\'anyi}}, \ and\
  \bibinfo {author} {\bibfnamefont {M.}~\bibnamefont {Ceriotti}},\ }\href
  {\doibase 10.1039/C6CP00415F} {\bibfield  {journal} {\bibinfo  {journal}
  {Phys. Chem. Chem. Phys.}\ }\textbf {\bibinfo {volume} {18}},\ \bibinfo
  {pages} {13754} (\bibinfo {year} {2016})}\BibitemShut {NoStop}%
\bibitem [{\citenamefont {M{\"u}ller}\ \emph {et~al.}(2001)\citenamefont
  {M{\"u}ller}, \citenamefont {Mika}, \citenamefont {R{\"a}tsch}, \citenamefont
  {Tsuda},\ and\ \citenamefont {Sch{\"o}lkopf}}]{muller2001introduction}%
  \BibitemOpen
  \bibfield  {author} {\bibinfo {author} {\bibfnamefont {K.-R.}\ \bibnamefont
  {M{\"u}ller}}, \bibinfo {author} {\bibfnamefont {S.}~\bibnamefont {Mika}},
  \bibinfo {author} {\bibfnamefont {G.}~\bibnamefont {R{\"a}tsch}}, \bibinfo
  {author} {\bibfnamefont {K.}~\bibnamefont {Tsuda}}, \ and\ \bibinfo {author}
  {\bibfnamefont {B.}~\bibnamefont {Sch{\"o}lkopf}},\ }\href@noop {} {\bibfield
   {journal} {\bibinfo  {journal} {IEEE transactions on neural networks}\
  }\textbf {\bibinfo {volume} {12}},\ \bibinfo {pages} {181} (\bibinfo {year}
  {2001})}\BibitemShut {NoStop}%
\bibitem [{\citenamefont {Sch{\"o}lkopf}\ and\ \citenamefont
  {Smola}(2002)}]{scholkopf2002learning}%
  \BibitemOpen
  \bibfield  {author} {\bibinfo {author} {\bibfnamefont {B.}~\bibnamefont
  {Sch{\"o}lkopf}}\ and\ \bibinfo {author} {\bibfnamefont {A.~J.}\ \bibnamefont
  {Smola}},\ }\href@noop {} {\emph {\bibinfo {title} {Learning with kernels:
  support vector machines, regularization, optimization, and beyond}}}\
  (\bibinfo  {publisher} {MIT press},\ \bibinfo {year} {2002})\BibitemShut
  {NoStop}%
\bibitem [{\citenamefont {Vovk}(2013)}]{Vovk2013}%
  \BibitemOpen
  \bibfield  {author} {\bibinfo {author} {\bibfnamefont {V.}~\bibnamefont
  {Vovk}},\ }\enquote {\bibinfo {title} {Kernel ridge regression},}\ in\ \href
  {\doibase 10.1007/978-3-642-41136-6_11} {\emph {\bibinfo {booktitle}
  {Empirical Inference: Festschrift in Honor of Vladimir N. Vapnik}}},\
  \bibinfo {editor} {edited by\ \bibinfo {editor} {\bibfnamefont
  {B.}~\bibnamefont {Sch{\"o}lkopf}}, \bibinfo {editor} {\bibfnamefont
  {Z.}~\bibnamefont {Luo}}, \ and\ \bibinfo {editor} {\bibfnamefont
  {V.}~\bibnamefont {Vovk}}}\ (\bibinfo  {publisher} {Springer Berlin
  Heidelberg},\ \bibinfo {address} {Berlin, Heidelberg},\ \bibinfo {year}
  {2013})\ pp.\ \bibinfo {pages} {105--116}\BibitemShut {NoStop}%
\bibitem [{\citenamefont {Hastie}\ \emph {et~al.}(2011)\citenamefont {Hastie},
  \citenamefont {Tibshirani},\ and\ \citenamefont
  {Friedman}}]{kernel_ridge_regression2}%
  \BibitemOpen
  \bibfield  {author} {\bibinfo {author} {\bibfnamefont {T.}~\bibnamefont
  {Hastie}}, \bibinfo {author} {\bibfnamefont {R.}~\bibnamefont {Tibshirani}},
  \ and\ \bibinfo {author} {\bibfnamefont {J.}~\bibnamefont {Friedman}},\
  }\href@noop {} {\emph {\bibinfo {title} {The Elements of Statistical
  Learning: Data Mining, Inference, and Prediction, Second Edition}}},\
  \bibinfo {edition} {2nd}\ ed.\ (\bibinfo  {publisher} {Springer},\ \bibinfo
  {address} {New York},\ \bibinfo {year} {2011})\BibitemShut {NoStop}%
\bibitem [{\citenamefont {Mathias}(2015)}]{mathias2015kernel}%
  \BibitemOpen
  \bibfield  {author} {\bibinfo {author} {\bibfnamefont {S.}~\bibnamefont
  {Mathias}},\ }\href@noop {} {\bibfield  {journal} {\bibinfo  {journal}
  {Master thesis:
  http://wissrech.ins.uni-bonn.de/teaching/master/masterthesis\_mathias\_revised.pdf}\
  } (\bibinfo {year} {2015})}\BibitemShut {NoStop}%
\bibitem [{\citenamefont {Barker}\ \emph {et~al.}(2016)\citenamefont {Barker},
  \citenamefont {Bulin}, \citenamefont {Hamaekers},\ and\ \citenamefont
  {Mathias}}]{barker2016localized}%
  \BibitemOpen
  \bibfield  {author} {\bibinfo {author} {\bibfnamefont {J.}~\bibnamefont
  {Barker}}, \bibinfo {author} {\bibfnamefont {J.}~\bibnamefont {Bulin}},
  \bibinfo {author} {\bibfnamefont {J.}~\bibnamefont {Hamaekers}}, \ and\
  \bibinfo {author} {\bibfnamefont {S.}~\bibnamefont {Mathias}},\ }\href@noop
  {} {\bibfield  {journal} {\bibinfo  {journal} {arXiv preprint
  arXiv:1611.05126}\ } (\bibinfo {year} {2016})}\BibitemShut {NoStop}%
\bibitem [{\citenamefont {Bartok}\ and\ \citenamefont
  {Csanyi}(2015)}]{Bartok2015GAP}%
  \BibitemOpen
  \bibfield  {author} {\bibinfo {author} {\bibfnamefont {A.~P.}\ \bibnamefont
  {Bartok}}\ and\ \bibinfo {author} {\bibfnamefont {G.}~\bibnamefont
  {Csanyi}},\ }\href {\doibase 10.1002/qua.24927} {\bibfield  {journal}
  {\bibinfo  {journal} {International Journal of Quantum Chemistry}\ }\textbf
  {\bibinfo {volume} {115}},\ \bibinfo {pages} {1051} (\bibinfo {year}
  {2015})}\BibitemShut {NoStop}%
\bibitem [{\citenamefont {Ramakrishnan}\ and\ \citenamefont {von
  Lilienfeld}(2015)}]{sk}%
  \BibitemOpen
  \bibfield  {author} {\bibinfo {author} {\bibfnamefont {R.}~\bibnamefont
  {Ramakrishnan}}\ and\ \bibinfo {author} {\bibfnamefont {O.~A.}\ \bibnamefont
  {von Lilienfeld}},\ }\href {\doibase doi:10.2533/chimia.2015.182} {\bibfield
  {journal} {\bibinfo  {journal} {chimia}\ }\textbf {\bibinfo {volume} {69}},\
  \bibinfo {pages} {182} (\bibinfo {year} {2015})}\BibitemShut {NoStop}%
\bibitem [{\citenamefont {Axilrod}\ and\ \citenamefont {Teller}(1943)}]{atm}%
  \BibitemOpen
  \bibfield  {author} {\bibinfo {author} {\bibfnamefont {B.~M.}\ \bibnamefont
  {Axilrod}}\ and\ \bibinfo {author} {\bibfnamefont {E.}~\bibnamefont
  {Teller}},\ }\href {\doibase 10.1063/1.1723844} {\bibfield  {journal}
  {\bibinfo  {journal} {J. Chem. Phys}\ }\textbf {\bibinfo {volume} {11}},\
  \bibinfo {pages} {299} (\bibinfo {year} {1943})}\BibitemShut {NoStop}%
\bibitem [{\citenamefont {Muto}(1943)}]{atm2}%
  \BibitemOpen
  \bibfield  {author} {\bibinfo {author} {\bibfnamefont {Y.}~\bibnamefont
  {Muto}},\ }\href@noop {} {\bibfield  {journal} {\bibinfo  {journal} {jpmsj}\
  }\textbf {\bibinfo {volume} {17}},\ \bibinfo {pages} {629} (\bibinfo {year}
  {1943})}\BibitemShut {NoStop}%
\bibitem [{\citenamefont {Christensen}\ \emph {et~al.}(2015)\citenamefont
  {Christensen}, \citenamefont {Elstner},\ and\ \citenamefont
  {Cui}}]{christensen2015improving}%
  \BibitemOpen
  \bibfield  {author} {\bibinfo {author} {\bibfnamefont {A.~S.}\ \bibnamefont
  {Christensen}}, \bibinfo {author} {\bibfnamefont {M.}~\bibnamefont
  {Elstner}}, \ and\ \bibinfo {author} {\bibfnamefont {Q.}~\bibnamefont
  {Cui}},\ }\href@noop {} {\bibfield  {journal} {\bibinfo  {journal} {The
  Journal of chemical physics}\ }\textbf {\bibinfo {volume} {143}},\ \bibinfo
  {pages} {084123} (\bibinfo {year} {2015})}\BibitemShut {NoStop}%
\bibitem [{\citenamefont {Ramakrishnan}\ \emph {et~al.}(2014)\citenamefont
  {Ramakrishnan}, \citenamefont {Dral}, \citenamefont {Rupp},\ and\
  \citenamefont {von Lilienfeld}}]{gdb9}%
  \BibitemOpen
  \bibfield  {author} {\bibinfo {author} {\bibfnamefont {R.}~\bibnamefont
  {Ramakrishnan}}, \bibinfo {author} {\bibfnamefont {P.~O.}\ \bibnamefont
  {Dral}}, \bibinfo {author} {\bibfnamefont {M.}~\bibnamefont {Rupp}}, \ and\
  \bibinfo {author} {\bibfnamefont {O.~A.}\ \bibnamefont {von Lilienfeld}},\
  }\href {http://dx.doi.org/10.1038/sdata.2014.22} {\bibfield  {journal}
  {\bibinfo  {journal} {Sci. Data}\ }\textbf {\bibinfo {volume} {1}} (\bibinfo
  {year} {2014})}\BibitemShut {NoStop}%
\bibitem [{\citenamefont {Stevens}\ \emph {et~al.}(1993)\citenamefont
  {Stevens}, \citenamefont {Devlin}, \citenamefont {Chabalowski},\ and\
  \citenamefont {Frisch}}]{B3LYP}%
  \BibitemOpen
  \bibfield  {author} {\bibinfo {author} {\bibfnamefont {P.~J.}\ \bibnamefont
  {Stevens}}, \bibinfo {author} {\bibfnamefont {F.~J.}\ \bibnamefont {Devlin}},
  \bibinfo {author} {\bibfnamefont {C.~F.}\ \bibnamefont {Chabalowski}}, \ and\
  \bibinfo {author} {\bibfnamefont {M.~J.}\ \bibnamefont {Frisch}},\
  }\href@noop {} {\bibfield  {journal} {\bibinfo  {journal} {J. Phys. Chem.}\
  }\textbf {\bibinfo {volume} {98}},\ \bibinfo {pages} {11623} (\bibinfo {year}
  {1993})}\BibitemShut {NoStop}%
\bibitem [{\citenamefont {Ruddigkeit}\ \emph {et~al.}(2012)\citenamefont
  {Ruddigkeit}, \citenamefont {van Deursen}, \citenamefont {Blum},\ and\
  \citenamefont {Reymond}}]{gdb17}%
  \BibitemOpen
  \bibfield  {author} {\bibinfo {author} {\bibfnamefont {L.}~\bibnamefont
  {Ruddigkeit}}, \bibinfo {author} {\bibfnamefont {R.}~\bibnamefont {van
  Deursen}}, \bibinfo {author} {\bibfnamefont {L.~C.}\ \bibnamefont {Blum}}, \
  and\ \bibinfo {author} {\bibfnamefont {J.-L.}\ \bibnamefont {Reymond}},\
  }\href {\doibase 10.1021/ci300415d} {\bibfield  {journal} {\bibinfo
  {journal} {J. Chem. Inf. Model.}\ }\textbf {\bibinfo {volume} {52}},\
  \bibinfo {pages} {2864} (\bibinfo {year} {2012})}\BibitemShut {NoStop}%
\bibitem [{\citenamefont {Montavon}\ \emph {et~al.}(2013)\citenamefont
  {Montavon}, \citenamefont {Rupp}, \citenamefont {Gobre}, \citenamefont
  {Vazquez-Mayagoitia}, \citenamefont {Hansen}, \citenamefont {Tkatchenko},
  \citenamefont {Müller},\ and\ \citenamefont {von
  Lilienfeld}}]{Montavon2013MLchemicalCompSpace}%
  \BibitemOpen
  \bibfield  {author} {\bibinfo {author} {\bibfnamefont {G.}~\bibnamefont
  {Montavon}}, \bibinfo {author} {\bibfnamefont {M.}~\bibnamefont {Rupp}},
  \bibinfo {author} {\bibfnamefont {V.}~\bibnamefont {Gobre}}, \bibinfo
  {author} {\bibfnamefont {A.}~\bibnamefont {Vazquez-Mayagoitia}}, \bibinfo
  {author} {\bibfnamefont {K.}~\bibnamefont {Hansen}}, \bibinfo {author}
  {\bibfnamefont {A.}~\bibnamefont {Tkatchenko}}, \bibinfo {author}
  {\bibfnamefont {K.-R.}\ \bibnamefont {Müller}}, \ and\ \bibinfo {author}
  {\bibfnamefont {O.~A.}\ \bibnamefont {von Lilienfeld}},\ }\href
  {http://stacks.iop.org/1367-2630/15/i=9/a=095003} {\bibfield  {journal}
  {\bibinfo  {journal} {New Journal of Physics}\ }\textbf {\bibinfo {volume}
  {15}},\ \bibinfo {pages} {095003} (\bibinfo {year} {2013})}\BibitemShut
  {NoStop}%
\bibitem [{\citenamefont {Blum}\ and\ \citenamefont {Reymond}(2009)}]{gdb13}%
  \BibitemOpen
  \bibfield  {author} {\bibinfo {author} {\bibfnamefont {L.~C.}\ \bibnamefont
  {Blum}}\ and\ \bibinfo {author} {\bibfnamefont {J.-L.}\ \bibnamefont
  {Reymond}},\ }\href@noop {} {\bibfield  {journal} {\bibinfo  {journal} {J.
  Am. Chem. Soc.}\ }\textbf {\bibinfo {volume} {131}},\ \bibinfo {pages} {8732}
  (\bibinfo {year} {2009})}\BibitemShut {NoStop}%
\bibitem [{\citenamefont {Ernzerhof}\ and\ \citenamefont
  {Scuseria}(1999)}]{PBE01}%
  \BibitemOpen
  \bibfield  {author} {\bibinfo {author} {\bibfnamefont {M.}~\bibnamefont
  {Ernzerhof}}\ and\ \bibinfo {author} {\bibfnamefont {G.~E.}\ \bibnamefont
  {Scuseria}},\ }\href@noop {} {\bibfield  {journal} {\bibinfo  {journal} {J.
  Chem. Phys.}\ }\textbf {\bibinfo {volume} {110}},\ \bibinfo {pages} {5029}
  (\bibinfo {year} {1999})}\BibitemShut {NoStop}%
\bibitem [{\citenamefont {Adamo}\ and\ \citenamefont {Barone}(1999)}]{PBE02}%
  \BibitemOpen
  \bibfield  {author} {\bibinfo {author} {\bibfnamefont {C.}~\bibnamefont
  {Adamo}}\ and\ \bibinfo {author} {\bibfnamefont {V.}~\bibnamefont {Barone}},\
  }\href@noop {} {\bibfield  {journal} {\bibinfo  {journal} {J. Chem. Phys.}\
  }\textbf {\bibinfo {volume} {110}},\ \bibinfo {pages} {6158} (\bibinfo {year}
  {1999})}\BibitemShut {NoStop}%
\bibitem [{\citenamefont {Burns}\ \emph {et~al.}(2017)\citenamefont {Burns},
  \citenamefont {Faver}, \citenamefont {Zheng}, \citenamefont {Marshall},
  \citenamefont {Smith}, \citenamefont {Vanommeslaeghe}, \citenamefont
  {MacKerellJr.}, \citenamefont {MerzJr.},\ and\ \citenamefont
  {Sherrill}}]{BurnsSSI}%
  \BibitemOpen
  \bibfield  {author} {\bibinfo {author} {\bibfnamefont {L.~A.}\ \bibnamefont
  {Burns}}, \bibinfo {author} {\bibfnamefont {J.~C.}\ \bibnamefont {Faver}},
  \bibinfo {author} {\bibfnamefont {Z.}~\bibnamefont {Zheng}}, \bibinfo
  {author} {\bibfnamefont {M.~S.}\ \bibnamefont {Marshall}}, \bibinfo {author}
  {\bibfnamefont {D.~G.~A.}\ \bibnamefont {Smith}}, \bibinfo {author}
  {\bibfnamefont {K.}~\bibnamefont {Vanommeslaeghe}}, \bibinfo {author}
  {\bibfnamefont {A.~D.}\ \bibnamefont {MacKerellJr.}}, \bibinfo {author}
  {\bibfnamefont {K.~M.}\ \bibnamefont {MerzJr.}}, \ and\ \bibinfo {author}
  {\bibfnamefont {C.~D.}\ \bibnamefont {Sherrill}},\ }\href {\doibase
  10.1063/1.5001028} {\bibfield  {journal} {\bibinfo  {journal} {The Journal of
  Chemical Physics}\ }\textbf {\bibinfo {volume} {147}},\ \bibinfo {pages}
  {161727} (\bibinfo {year} {2017})}\BibitemShut {NoStop}%
\bibitem [{\citenamefont {Marshall}\ and\ \citenamefont
  {Sherrill}(2011)}]{Marshall2011}%
  \BibitemOpen
  \bibfield  {author} {\bibinfo {author} {\bibfnamefont {M.~S.}\ \bibnamefont
  {Marshall}}\ and\ \bibinfo {author} {\bibfnamefont {C.~D.}\ \bibnamefont
  {Sherrill}},\ }\href@noop {} {\bibfield  {journal} {\bibinfo  {journal}
  {Journal of Chemical Theory and Computation}\ }\textbf {\bibinfo {volume}
  {7}},\ \bibinfo {pages} {3978} (\bibinfo {year} {2011})}\BibitemShut
  {NoStop}%
\bibitem [{\citenamefont {Jorgensen}\ \emph {et~al.}(1983)\citenamefont
  {Jorgensen}, \citenamefont {Chandrasekhar}, \citenamefont {Madura},
  \citenamefont {Impey},\ and\ \citenamefont {Klein}}]{Tip3p}%
  \BibitemOpen
  \bibfield  {author} {\bibinfo {author} {\bibfnamefont {W.~L.}\ \bibnamefont
  {Jorgensen}}, \bibinfo {author} {\bibfnamefont {J.}~\bibnamefont
  {Chandrasekhar}}, \bibinfo {author} {\bibfnamefont {J.~D.}\ \bibnamefont
  {Madura}}, \bibinfo {author} {\bibfnamefont {R.~W.}\ \bibnamefont {Impey}}, \
  and\ \bibinfo {author} {\bibfnamefont {M.~L.}\ \bibnamefont {Klein}},\
  }\href@noop {} {\bibfield  {journal} {\bibinfo  {journal} {The Journal of
  chemical physics}\ }\textbf {\bibinfo {volume} {79}},\ \bibinfo {pages} {926}
  (\bibinfo {year} {1983})}\BibitemShut {NoStop}%
\bibitem [{\citenamefont {Brooks}\ \emph {et~al.}(1983)\citenamefont {Brooks},
  \citenamefont {Bruccoleri}, \citenamefont {Olafson}, \citenamefont {States},
  \citenamefont {Swaminathan},\ and\ \citenamefont
  {Karplus}}]{brooks1983charmm}%
  \BibitemOpen
  \bibfield  {author} {\bibinfo {author} {\bibfnamefont {B.~R.}\ \bibnamefont
  {Brooks}}, \bibinfo {author} {\bibfnamefont {R.~E.}\ \bibnamefont
  {Bruccoleri}}, \bibinfo {author} {\bibfnamefont {B.~D.}\ \bibnamefont
  {Olafson}}, \bibinfo {author} {\bibfnamefont {D.~J.}\ \bibnamefont {States}},
  \bibinfo {author} {\bibfnamefont {S.~a.}\ \bibnamefont {Swaminathan}}, \ and\
  \bibinfo {author} {\bibfnamefont {M.}~\bibnamefont {Karplus}},\ }\href@noop
  {} {\bibfield  {journal} {\bibinfo  {journal} {Journal of computational
  chemistry}\ }\textbf {\bibinfo {volume} {4}},\ \bibinfo {pages} {187}
  (\bibinfo {year} {1983})}\BibitemShut {NoStop}%
\bibitem [{\citenamefont {Grimme}\ \emph {et~al.}(2015)\citenamefont {Grimme},
  \citenamefont {Brandenburg}, \citenamefont {Bannwarth},\ and\ \citenamefont
  {Hansen}}]{grimme2015consistent}%
  \BibitemOpen
  \bibfield  {author} {\bibinfo {author} {\bibfnamefont {S.}~\bibnamefont
  {Grimme}}, \bibinfo {author} {\bibfnamefont {J.~G.}\ \bibnamefont
  {Brandenburg}}, \bibinfo {author} {\bibfnamefont {C.}~\bibnamefont
  {Bannwarth}}, \ and\ \bibinfo {author} {\bibfnamefont {A.}~\bibnamefont
  {Hansen}},\ }\href@noop {} {\bibfield  {journal} {\bibinfo  {journal} {The
  Journal of chemical physics}\ }\textbf {\bibinfo {volume} {143}},\ \bibinfo
  {pages} {054107} (\bibinfo {year} {2015})}\BibitemShut {NoStop}%
\bibitem [{\citenamefont {Neese}(2012)}]{neese2012orca}%
  \BibitemOpen
  \bibfield  {author} {\bibinfo {author} {\bibfnamefont {F.}~\bibnamefont
  {Neese}},\ }\href@noop {} {\bibfield  {journal} {\bibinfo  {journal} {Wiley
  Interdisciplinary Reviews: Computational Molecular Science}\ }\textbf
  {\bibinfo {volume} {2}},\ \bibinfo {pages} {73} (\bibinfo {year}
  {2012})}\BibitemShut {NoStop}%
\bibitem [{\citenamefont {Belsky}\ \emph {et~al.}(2002)\citenamefont {Belsky},
  \citenamefont {Hellenbrandt}, \citenamefont {Karen},\ and\ \citenamefont
  {Luksch}}]{icsd}%
  \BibitemOpen
  \bibfield  {author} {\bibinfo {author} {\bibfnamefont {A.}~\bibnamefont
  {Belsky}}, \bibinfo {author} {\bibfnamefont {M.}~\bibnamefont
  {Hellenbrandt}}, \bibinfo {author} {\bibfnamefont {V.~L.}\ \bibnamefont
  {Karen}}, \ and\ \bibinfo {author} {\bibfnamefont {P.}~\bibnamefont
  {Luksch}},\ }\href {\doibase 10.1107/S0108768102006948} {\bibfield  {journal}
  {\bibinfo  {journal} {Acta Crystallographica Section B Structural Science}\
  }\textbf {\bibinfo {volume} {58}},\ \bibinfo {pages} {364} (\bibinfo {year}
  {2002})}\BibitemShut {NoStop}%
\bibitem [{\citenamefont {Bergerhoff}\ \emph {et~al.}(1983)\citenamefont
  {Bergerhoff}, \citenamefont {Hundt}, \citenamefont {Sievers},\ and\
  \citenamefont {Brown}}]{icsd2}%
  \BibitemOpen
  \bibfield  {author} {\bibinfo {author} {\bibfnamefont {G.}~\bibnamefont
  {Bergerhoff}}, \bibinfo {author} {\bibfnamefont {R.}~\bibnamefont {Hundt}},
  \bibinfo {author} {\bibfnamefont {R.}~\bibnamefont {Sievers}}, \ and\
  \bibinfo {author} {\bibfnamefont {I.~D.}\ \bibnamefont {Brown}},\ }\href
  {\doibase 10.1021/ci00038a003} {\bibfield  {journal} {\bibinfo  {journal}
  {Journal of Chemical Information and Computer Sciences}\ }\textbf {\bibinfo
  {volume} {23}},\ \bibinfo {pages} {66} (\bibinfo {year} {1983})}\BibitemShut
  {NoStop}%
\bibitem [{\citenamefont {Kirklin}\ \emph {et~al.}(2015)\citenamefont
  {Kirklin}, \citenamefont {Saal}, \citenamefont {Meredig}, \citenamefont
  {Thompson}, \citenamefont {Doak}, \citenamefont {Aykol}, \citenamefont
  {R{\"u}hl},\ and\ \citenamefont {Wolverton}}]{kirklin2015open}%
  \BibitemOpen
  \bibfield  {author} {\bibinfo {author} {\bibfnamefont {S.}~\bibnamefont
  {Kirklin}}, \bibinfo {author} {\bibfnamefont {J.~E.}\ \bibnamefont {Saal}},
  \bibinfo {author} {\bibfnamefont {B.}~\bibnamefont {Meredig}}, \bibinfo
  {author} {\bibfnamefont {A.}~\bibnamefont {Thompson}}, \bibinfo {author}
  {\bibfnamefont {J.~W.}\ \bibnamefont {Doak}}, \bibinfo {author}
  {\bibfnamefont {M.}~\bibnamefont {Aykol}}, \bibinfo {author} {\bibfnamefont
  {S.}~\bibnamefont {R{\"u}hl}}, \ and\ \bibinfo {author} {\bibfnamefont
  {C.}~\bibnamefont {Wolverton}},\ }\href@noop {} {\bibfield  {journal}
  {\bibinfo  {journal} {npj Computational Materials}\ }\textbf {\bibinfo
  {volume} {1}},\ \bibinfo {pages} {15010} (\bibinfo {year}
  {2015})}\BibitemShut {NoStop}%
\bibitem [{\citenamefont {Saal}\ \emph {et~al.}(2013)\citenamefont {Saal},
  \citenamefont {Kirklin}, \citenamefont {Aykol}, \citenamefont {Meredig},\
  and\ \citenamefont {Wolverton}}]{Saal2013}%
  \BibitemOpen
  \bibfield  {author} {\bibinfo {author} {\bibfnamefont {J.~E.}\ \bibnamefont
  {Saal}}, \bibinfo {author} {\bibfnamefont {S.}~\bibnamefont {Kirklin}},
  \bibinfo {author} {\bibfnamefont {M.}~\bibnamefont {Aykol}}, \bibinfo
  {author} {\bibfnamefont {B.}~\bibnamefont {Meredig}}, \ and\ \bibinfo
  {author} {\bibfnamefont {C.}~\bibnamefont {Wolverton}},\ }\href {\doibase
  10.1007/s11837-013-0755-4} {\bibfield  {journal} {\bibinfo  {journal} {JOM}\
  }\textbf {\bibinfo {volume} {65}},\ \bibinfo {pages} {1501} (\bibinfo {year}
  {2013})}\BibitemShut {NoStop}%
\bibitem [{\citenamefont {Ward}\ \emph {et~al.}(2017)\citenamefont {Ward},
  \citenamefont {Liu}, \citenamefont {Krishna}, \citenamefont {Hegde},
  \citenamefont {Agrawal}, \citenamefont {Choudhary},\ and\ \citenamefont
  {Wolverton}}]{Ward2017Voronoi}%
  \BibitemOpen
  \bibfield  {author} {\bibinfo {author} {\bibfnamefont {L.}~\bibnamefont
  {Ward}}, \bibinfo {author} {\bibfnamefont {R.}~\bibnamefont {Liu}}, \bibinfo
  {author} {\bibfnamefont {A.}~\bibnamefont {Krishna}}, \bibinfo {author}
  {\bibfnamefont {V.~I.}\ \bibnamefont {Hegde}}, \bibinfo {author}
  {\bibfnamefont {A.}~\bibnamefont {Agrawal}}, \bibinfo {author} {\bibfnamefont
  {A.}~\bibnamefont {Choudhary}}, \ and\ \bibinfo {author} {\bibfnamefont
  {C.}~\bibnamefont {Wolverton}},\ }\href {\doibase 10.1103/PhysRevB.96.024104}
  {\bibfield  {journal} {\bibinfo  {journal} {Phys. Rev. B}\ }\textbf {\bibinfo
  {volume} {96}},\ \bibinfo {pages} {024104} (\bibinfo {year}
  {2017})}\BibitemShut {NoStop}%
\bibitem [{\citenamefont {Perdew}\ \emph {et~al.}(1996)\citenamefont {Perdew},
  \citenamefont {Burke},\ and\ \citenamefont {Ernzerhof}}]{PBE}%
  \BibitemOpen
  \bibfield  {author} {\bibinfo {author} {\bibfnamefont {J.~P.}\ \bibnamefont
  {Perdew}}, \bibinfo {author} {\bibfnamefont {K.}~\bibnamefont {Burke}}, \
  and\ \bibinfo {author} {\bibfnamefont {M.}~\bibnamefont {Ernzerhof}},\
  }\href@noop {} {\bibfield  {journal} {\bibinfo  {journal} {Phys. Rev. Lett.}\
  }\textbf {\bibinfo {volume} {77}},\ \bibinfo {pages} {3865} (\bibinfo {year}
  {1996})}\BibitemShut {NoStop}%
\bibitem [{\citenamefont {Ramakrishnan}\ \emph {et~al.}(2015)\citenamefont
  {Ramakrishnan}, \citenamefont {Dral}, \citenamefont {Rupp},\ and\
  \citenamefont {von Lilienfeld}}]{deltalearning}%
  \BibitemOpen
  \bibfield  {author} {\bibinfo {author} {\bibfnamefont {R.}~\bibnamefont
  {Ramakrishnan}}, \bibinfo {author} {\bibfnamefont {P.~O.}\ \bibnamefont
  {Dral}}, \bibinfo {author} {\bibfnamefont {M.}~\bibnamefont {Rupp}}, \ and\
  \bibinfo {author} {\bibfnamefont {O.~A.}\ \bibnamefont {von Lilienfeld}},\
  }\href {\doibase 10.1021/acs.jctc.5b00099} {\bibfield  {journal} {\bibinfo
  {journal} {J. Chem. Theory Comput.}\ }\textbf {\bibinfo {volume} {11}},\
  \bibinfo {pages} {2087} (\bibinfo {year} {2015})}\BibitemShut {NoStop}%
\bibitem [{\citenamefont {Bartok}\ \emph {et~al.}(2017)\citenamefont {Bartok},
  \citenamefont {De}, \citenamefont {Poelking}, \citenamefont {Bernstein},
  \citenamefont {Kermode}, \citenamefont {Csanyi},\ and\ \citenamefont
  {Ceriotti}}]{bartok2017machine}%
  \BibitemOpen
  \bibfield  {author} {\bibinfo {author} {\bibfnamefont {A.~P.}\ \bibnamefont
  {Bartok}}, \bibinfo {author} {\bibfnamefont {S.}~\bibnamefont {De}}, \bibinfo
  {author} {\bibfnamefont {C.}~\bibnamefont {Poelking}}, \bibinfo {author}
  {\bibfnamefont {N.}~\bibnamefont {Bernstein}}, \bibinfo {author}
  {\bibfnamefont {J.}~\bibnamefont {Kermode}}, \bibinfo {author} {\bibfnamefont
  {G.}~\bibnamefont {Csanyi}}, \ and\ \bibinfo {author} {\bibfnamefont
  {M.}~\bibnamefont {Ceriotti}},\ }\href@noop {} {\bibfield  {journal}
  {\bibinfo  {journal} {arXiv preprint arXiv:1706.00179}\ } (\bibinfo {year}
  {2017})}\BibitemShut {NoStop}%
\bibitem [{\citenamefont {Bart\'ok}\ \emph {et~al.}(2013)\citenamefont
  {Bart\'ok}, \citenamefont {Kondor},\ and\ \citenamefont
  {Cs\'anyi}}]{SOAP_original}%
  \BibitemOpen
  \bibfield  {author} {\bibinfo {author} {\bibfnamefont {A.~P.}\ \bibnamefont
  {Bart\'ok}}, \bibinfo {author} {\bibfnamefont {R.}~\bibnamefont {Kondor}}, \
  and\ \bibinfo {author} {\bibfnamefont {G.}~\bibnamefont {Cs\'anyi}},\ }\href
  {\doibase 10.1103/PhysRevB.87.184115} {\bibfield  {journal} {\bibinfo
  {journal} {Phys. Rev. B}\ }\textbf {\bibinfo {volume} {87}},\ \bibinfo
  {pages} {184115} (\bibinfo {year} {2013})}\BibitemShut {NoStop}%
\bibitem [{\citenamefont {Faber}\ \emph {et~al.}(2015)\citenamefont {Faber},
  \citenamefont {Lindmaa}, \citenamefont {von Lilienfeld},\ and\ \citenamefont
  {Armiento}}]{qc_Felix_crystal}%
  \BibitemOpen
  \bibfield  {author} {\bibinfo {author} {\bibfnamefont {F.}~\bibnamefont
  {Faber}}, \bibinfo {author} {\bibfnamefont {A.}~\bibnamefont {Lindmaa}},
  \bibinfo {author} {\bibfnamefont {O.~A.}\ \bibnamefont {von Lilienfeld}}, \
  and\ \bibinfo {author} {\bibfnamefont {R.}~\bibnamefont {Armiento}},\ }\href
  {\doibase 10.1002/qua.24917} {\bibfield  {journal} {\bibinfo  {journal} {Int.
  J. Quantum}\ }\textbf {\bibinfo {volume} {115}},\ \bibinfo {pages} {1094}
  (\bibinfo {year} {2015})}\BibitemShut {NoStop}%
\bibitem [{\citenamefont {von Lilienfeld}(2009)}]{anatole-jcp2009-2}%
  \BibitemOpen
  \bibfield  {author} {\bibinfo {author} {\bibfnamefont {O.~A.}\ \bibnamefont
  {von Lilienfeld}},\ }\href@noop {} {\bibfield  {journal} {\bibinfo  {journal}
  {J. Chem. Phys.}\ }\textbf {\bibinfo {volume} {131}},\ \bibinfo {pages}
  {164102} (\bibinfo {year} {2009})}\BibitemShut {NoStop}%
\end{thebibliography}%
\end{document}